\documentclass[]{raa}
\usepackage{tipa}
\usepackage{bbm}
\usepackage{mathrsfs}
\usepackage{txfonts}
\usepackage{amssymb}            
\usepackage{graphicx,times}
\usepackage{natbib}

\usepackage{lscape}
\usepackage{textcomp}
\usepackage{epsfig}

\providecommand{\bm}[1]{\mbox{\boldmath$#1$\unboldmath}}

\begin{document}

   \title{HNCO: A Molecule Traces Low-velocity Shock
}

 \volnopage{ {\bf 20xx} Vol.\ {\bf 9} No. {\bf XX}, 000--000}
   \setcounter{page}{1}

   \author{Naiping Yu\inst{1,2}, Jinglong Xu\inst{1,2}, Jun-Jie Wang\inst{1,2}
   }

   \institute{National Astronomical Observatories, Chinese Academy of Sciences,
             Beijing 100012, China; {\it yunaiping09@mails.gucas.ac.cn}\\
        \and
             NAOC-TU Joint Center for Astrophysics, Lhasa 850000, China
        \and
\vs \no
   {\small Received [year] [month] [day]; accepted [year] [month] [day]
} }

\abstract{Using data from MALT90 (Millimetre Astronomy Legacy Team Survey at
90 GHz), we present molecular line study of a sample of ATLASGAL
(APEX Telescope Large Area Survey of the Galaxy) clumps. Twelve
emission lines have been detected in all. We found that in most
sources, emissions of HC$_3$N, HN$^{13}$C, CH$_3$CN, HNCO and SiO
show more compact distributions than those of HCO$^+$, HNC,
HCN and N$_2$H$^+$. By comparing with other molecular lines, we
found that the abundance of HNCO ($\chi$(HNCO)) correlates well with
other species such as HC$_3$N, HNC, C$_2$H,  H$^{13}$CO$^+$ and N$_2$H$^+$.
Previous studies indicate the HNCO abundance could be enhanced
by shocks. However, in this study, we found the abundance of HNCO
does not correlate well with that of SiO, which is also a good
tracer of shocks. We suggest this may be because HNCO and SiO trace different
parts of shocks. Our analysis indicates that the velocity of shock traced by HNCO tends to be lower than that traced by SiO. In the low-velocity
shocks traced by HNCO, the HNCO abundance increases faster than that
of SiO. While in the relatively high-velocity shocks traced by SiO, the SiO
abundance increases faster than that of HNCO. We suggest that in the
infrared dark cloud (IRDC) of MSXDC G331.71+00.59, high-velocity
shocks are destroying the molecule of HNCO.  }

   \maketitle


%
%
\section{Introduction}
The emission of interstellar isocyanic acid (HNCO) was first
detected by Snyder $\&$ Buhl (1972) in Sgr B2. Since its discovery,
HNCO has been found to be ubiquitous in our Galactic star-forming
regions (e.g. Brown 1981; Churchwell et al. 1986; Zinchenko et al.
2000). It has also been detected in extra-galaxies and the
circumstellar envelopes around asymptotic giant branch (AGB) stars
(e.g. Meier $\&$ Turner 2005;  Prieto et al. 2015). Observations
indicate HNCO traces the densest part of molecular clouds (Jackson
et al. 1984). Zinchenko et al . (2000) suggest the HNCO abundance
could be enhanced by shocked gas. They found the HNCO integrated
line intensities correlate well with those of SiO emissions
in massive galactic dense cores, indicating a similar production
mechanism of the two species. Mart\'in  et al. (2008) conducted a
multitransition study toward 13 molecular clouds in the Galactic
center region. They found the HNCO/$^{13}$CS relative abundance
ratio was very sensitive to UV radiations and shocks. This abundance
ratio could be used as a tool to distinguish between the
influence of shocks and the radiation field in molecular clouds. Li
et al. (2013) mapped 9 massive star-forming regions of HNCO emissions with the
Purple Mountain Observatory (PMO) 13.7 m telescope. They found possible shock enhancement of HNCO in Orion KL and W75OH. indicating shocks could enhance
the HNCO abundance. They also found that the line
parameters of HNCO and HC$_3$N have good correlations, implying
similar excitation mechanisms for the two species.

Early work suggests HNCO mainly be formed through gas-phase
reactions (e.g. Iglesias 1977; Turner et al. 1999). However,
chemical models reveal that the abundances of HNCO derived from
gas-phase reactions are inconsistent with observations ( e.g.
Tideswell et al. 2010). To explain the observed abundances of HNCO
in star-forming regions, reactions on the surfaces of grains should
be involved. In the early stage of star formation, as material collapses, a wide array of atoms and
molecules are absorbed onto the dust grains. Grain-surface chemistry
becomes available. Calculations indicate that it is an effective way to
produce HNCO on grain mantles (e.g. Allen $\&$ Robinson 1977; Garrod
et al. 2008). When the gas in a region is shocked, HNCO could be
ejected into the gas-phase by low-velocity shocks (e.g. Flower et al. 1995; Mart\'{i}n et al. 2008). In the hot core
stage of star formation, the HNCO emissions we observed could also
be formed by the destruction of more complex molecules (such as
HNCHO, HNCOCHO, HNCONH and HNCOOH) after their evaporation into the
gas phase (e.g. Tideswell et al. 2010). The destruction of HNCO in
star-forming regions are dominated by reactions with H$_e$$^+$ and
H$_3$$^+$ (Turner et al. 1999). It can also be destroyed by FUV
photons and cosmic rays.

In order to investigate the physical and chemical properties of HNCO
in star-forming regions, we present a molecular line study of HNCO
in a sample of ATLASGAL clumps. We introduce our data and source
selections in Section 2, results in Section 3, analysis in Section
4, and summary in Section 5.

\section{Data}
Our molecular line data comes from MALT90, which is carried out by
the Mopra 22-m telescope. MALT90 is an international project aimed at
characterizing the physical and chemical evolution of massive
star-forming regions within our Galaxy (e.g. Foster et al. 2011;
Jackson et al. 2013). Mopra is equipped with three receivers for
single-dish observations. Near 90 GHz, the angular resolution of
Mopra is 38$^{\prime\prime}$, and the antenna efficiency ranges from
0.49 at 86 GHz to 0.42 at 115 GHz (Ladd et al. 2005). MALT90 maps 16
emission lines simultaneously. The velocity resolution of the data
is about 0.11 km s$^{-1}$, with pointing accuracy $\sim$
8$^{\prime\prime}$, and the absolute flux uncertainty ranges from 10
to 17$\%$ depending on the transition in question (Foster et al.
2013). The target clumps of this survey are selected from the 870
$\mu$m sky survey of ATLASGAL (Schuller et al. 2009; Contreras et
al. 2013). The data files are publicly available and can be
downloaded from the MALT90 Home Page\footnote{
http://atoa.atnf.csiro.au/MALT90}. Using software packages of CLASS
(Continuum and Line Analysis Single-Disk Software) and GREG
(Grenoble Graphic), we conducted the data. MALT90 observed more than 3000 ATLASGAL clumps. Among the 16
emission lines, HNCO (4$_{0,4}$ - 3$_{0,3}$) is one of the least
detected ones (Rathborne et al. 2016). In order to study HNCO, we
searched the brightest 300 ATLASGAL clumps and selected 18 sources
which show distinct HNCO emissions. Sources in the Galactic center
are not included because sources in that direction used to have
multiple velocity components. The basic information of our sources
is listed in Table 1.

Dust temperature (T$_d$) is essential in the study of chemical
evolution in star formation regions. When T$_d$ is below $\sim$20 K,
carbon species like CO and CS can be depleted in the cold gas.
Derived from adjusting single-temperature dust emission
models to the far-infrared intensity maps measured between
160 and 870 $\mu$m from the Herschel and APEX sky surveys,
Guzm\'{a}n et al. (2015) recently have calculated the dust
temperatures and H$_2$ column densities for $\sim$3000 MALT90
clumps. The FWHM of these data differs from 12$^\prime$$^\prime$ to 35$^\prime$$^\prime$ (the
data from the PACS 70 $\mu$m band are excluded). To make an
adequate comparison, they convolved all images to a spatial
resolution of 35$^\prime$$^\prime$, which is the lowest resolution given by the
500 $\mu$m SPIRE instrument. We use these dust temperatures
and H$_2$ column densities derived by Guzm\'{a}n et al. (2015) in
our study due to the similar beam resolution of Mopra (35$^\prime$$^\prime$
versus 38$^\prime$$^\prime$). These values are also listed in Table 1. The mean dust temperature of our sample is 23.9 K.
This is consistent with many other observations of MYSOs and
H II regions (e.g., Hennemann et al. 2009; Sreenilayam $\&$
Fich 2011).

\section{Results}
Among the 16 spectral lines,  HCO$^+$ (1-0), HCN (1-0), HNC (1-0),
N$_2$H$^+$ (1-0), C$_2$H (1-0), H$^{13}$CO$^+$ (1-0), HC$_3$N
(10-9), HN$^{13}$C (1-0), HNCO (4$_{0,4}$ - 3$_{0,3}$) and SiO (2-1)
are detected in all sources. While $^{13}$C$^{34}$S (2-1), HNCO
(4$_{1,3}$ - 3$_{1,2}$), HC$^{13}$CCN (10-9) and H41$\alpha$ are
detected in none. Fig.1 to Fig.18 show the detected lines and their
integrated intensities. It can be noted that emissions of HC$_3$N,
HN$^{13}$C, HNCO and SiO are more compact than those of
HCO$^+$, HCN, HNC and N$_2$H$^+$ in most sources. We characterize the size of a molecular cloud by using the beam deconvolved angular diameter of a circle with
the same area as the half peak intensity:
\begin{equation}
\theta (species) = 2 (\frac{A_{1/2}}{\pi} - \frac{\theta^2_{beam}}{4})^{1/2}
\end{equation}
where $A_{1/2}$ is the area within the contour of half peak intensity, $\theta_{beam}$ the FWHM beam size (38$^\prime$$^\prime$). The beam deconvolved angular diameters of different species are presented in Table 2. In Fig 19, we present plots that compare beam deconvolved angular diameters of HNCO and those of other species. The sizes of HNCO clumps are comparable to those of SiO clumps.

In many cases, the HCO$^+$ (1-0) and HNC (1-0) lines show wide wing emissions and so-called ``blue profile", indicating outflow and
infall activities in large scale. Near 90 GHz, N$_2$H$^+$ (1-0) has seven
hyperfine transitions. These transitions blend into three groups
because of turbulence. The HCN (1-0) rotational transition split
into three hyperfine structures. However, as are shown in the
figures, these components always exhibit extended wing emissions, and
their self-absorbed line profiles used to be blended, preventing us
from doing following analysis. The H$^{13}$CO$^+$ (1-0) and $^{13}$CS (1-0), especially $^{13}$CS (1-0),
seems to be depleted in several sources. In the
survey of MALT90, there are five CH$_3$CN (5-4) hyperfine transitions.
However, the CH$_3$CN (5$_0$-4$_0$) and (5$_1$-4$_1$) two components
use to be blended. Only in the source of AGAL327.293-00.579 all the
five components are distinctly detected.

Assuming local thermodynamic equilibrium (LTE) conditions and a beam
filling factor of 1, we estimate column densities of HNCO, HC$_3$N,
N$_2$H$^+$, HNC, $^{13}$CS, C$_2$H, H$^{13}$CO$^+$ and SiO by using
equation
\begin{equation}
N(species) = \frac{8 \pi \nu^3}{c^3 R} \frac{Q_{rot}}{g_u A_{ul}}
\frac{exp(E_l/k T_{ex})}{1 - exp (-h \nu /k T_{ex})} \frac{\tau}{1 -
exp(-\tau)} \frac{\int T_{mb} dv}{J(T_{ex}) - J(T_{bg})}
\end{equation}
where $c$ is the velocity of light in the vacuum, $\nu$ is the
frequency of the transitions, $g_u$ is the statistical weight of the
upper level, $A_{ul}$ is the Einstein coefficient, $E_l$ is the
energy of the lower level, $Q_{rot}$ is the partition function,
$\tau$ is the optical depth, $T_{bg}$ (2.73 K) and $T_{ex}$ are the
temperature of the background radiation and the excitation
temperature in all cases. We assume that $T_{ex}$ is equal to the
dust temperature ($T_d$) derived by Guzm\'{a}n et al. (2015) (Table 1). $R$ is
only relevant for hyperfine transitions because it takes into
account the satellite lines correcting by their relative opacities.
The value of $R$ is 5/9 for N$_2$H$^+$, 5/12 for C$_2$H and 1.0 for
transitions without hyperfine structure. The values of $g_u$,
$A_{ul}$ and $E_l$ could be found in the Cologne Database for
Molecular Spectroscopy (CDMS) (M\"{u}ller et al. 2001, 2005 ).
$J(T)$ is defined by
\begin{equation}
J(T) = \frac{h \nu}{k} \frac{1}{e^{h \nu/k T} - 1}
\end{equation}

The partition functions ($Q_{rot}$) of the linear, rigid rotor
molecules (N$_2$H$^+$, H$^{13}$CO$^+$, HNC and SiO) could be approximated
as
\begin{equation}
Q_{rot} \simeq \frac{k T_{ex}}{h B} + \frac{1}{3}
\end{equation}
where $B$ is the rotational constant. The partition functions of
C$_2$H, HC$_3$N and HNCO can be found in Table 7 of Sanhueza et al.
(2012).

We assume emissions of $^{13}$CS, HC$_3$N, SiO and HNCO are
optically thin. This assumption is suitable as shapes of these line
are relatively simple in all sources. To derive the line parameters,
we fit these line emissions with a single Gaussian profile from the
averaged pixels inside 38$^{\prime\prime}$. The line widths, peak
emissions and integrated intensities are listed in Table 3, Table 4
and Table 5 respectively. The derived column densities are listed in Table 6.

For HCO$^+$ and HNC, the presence of their isotopoloques allows us
to estimate their optical thickness through
\begin{equation}
\frac{1 - e^{-\tau_{12}}}{1 - e^{-\tau_{12}/X}} =
\frac{^{12}T_{mb}}{^{13}T_{mb}}
\end{equation}
where $X$ $\sim$ $[^{12}C]/[^{13}C]$ is the isotope abundance ratio.
Here we use a constant $X$ = 50 in our calculations (Purcell et al.
2006).

In the case of N$_2$H$^+$, we follow the procedure described by
Purcell et al. (2009) to estimate the optical depth. Assuming the
line widths of the individual hyperfine components of N$_2$H$^+$ are
all equal and optically thin, the integrated intensities of group
1/group 2 (defined by Purcell et al. 2009) should be in the ratio of
1:5. The optical depth can then be derived from the ratio of the two
integrated intensities, using the following equation:
\begin{equation}
\frac{\int T_{MB,1} dv}{\int T_{MB,2} dv} = \frac{1 -
exp(-0.2\tau_2)}{1 - exp(-\tau_2)}
\end{equation}
To derive the line intensities and peak emissions, we fit the
three groups with three Gaussian profiles. The integrated intensities of group 2
are listed in table 5 and the derived column densities of N$_2$H$^+$
are presented in table 6.

Near 90 GHz, C$_2$H ($N = 1 - 0$) splits into six hyperfine
transitions out of which two ($N = 1 - 0, J = 3/2 - 1/2, F = 2 - 1$
and $N = 1 - 0, J = 3/2 - 1/2, F = 1 - 0$) could easily be detected.
The optical depth of C$_2$H ($F=2-1$) can then be derived by
comparing with its hyperfine components. In the case of optically
thin limit, the intensity ratio of C$_2$H ($F=2-1$) and C$_2$H
($F=1-0$) should be 2.0 (Tucker et al. 1974). Thus, the opacity of
C$_2$H ($F=2-1$) could be given by
\begin{equation}
\frac{1 - e^{-0.5\tau}}{1 - e^{-\tau}} =
\frac{T_{mb}(F=1-0)}{T_{mb}(F=2-1)}
\end{equation}
The column densities of C$_2$H can then be calculated through
equation 1. The calculated values are
listed in table 6.

To transfer column densities to abundances, we use
the $N(species)/N(H_2)$ ratio, where $N(H_2)$ values were estimated by Guzm\'{a}n et al. (2015) (Table 1). The derived abundances are presented in Table 7.

\section{Analysis}
Fig. 20 shows correlation plots of the HNCO abundance against those
of other species. Table 8 presents the least-square fitting functions and coefficients for the correlations between HNCO and these species.
It can be noted that in our sample, the abundances
of HC$_3$N, HNC, N$_2$H$^+$, C$_2$H and H$^{13}$CO$^+$ have a
good correlation with $\chi$(HNCO). While the correlations between
$\chi$(HNCO) and $\chi$($^{13}$CS), $\chi$(SiO) are not
so good. We should mention here that the beam size of Mopra is 38$^\prime$$^\prime$, which at a distance
of 3 kpc for high-mass star-forming regions, has a physical size
of $\sim$ 0.6 pc. Therefore, our data probes star forming clumps,
which may contain several star-forming cores (size $<$ 0.1 pc),
and also diffuse material.

HC$_3$N is regarded as a good tracer of warm dense gas (e.g.
Miettinen 2014). Using the PMO 13.7 m
telescope, Li et al. (2013) studied spatial distributions of HNCO
in nine massive star-forming regions. They found the integrated
intensities, line widths and LSR velocities of HC$_3$N and HNCO
correlate well with each other. On the other hand, also based on the
data of MALT90, Miettinen (2014) found that in some infrared dark
clouds, the emission morphology of HC$_3$N resembles those of HNCO,
HNC and N$_2$H$^+$. These studies imply HC$_3$N has similar
excitation mechanism with HNCO. As shown in the top left panel of
Fig. 20, there is a hint that the abundance of HC$_3$N  increases as
a function of $\chi$(HNCO). The functional form of the linear fit is
$\chi$(HC$_3$N) = 0.89 + 0.24 $\chi$(HNCO). Previous studies
indicate HC$_3$N is mainly produced through C$_2$H$_2$ + CN
$\rightarrow$ HC$_3$N + H in star-forming regions (Chapman et al.
2009). CN is also the main material to produce HNCO. In shocked gas,
CN can form OCN through CN + O$_2$ $\rightarrow$ OCN + O (Turner et
al. 1999). OCN can further produce HNCO both through gas-phase (OCN
+ H$_2$ $\rightarrow$ HNCO + H) and/or grain surface reactions (OCN + H
$\rightarrow$ HNCO) (e.g. Allen $\&$ Robinson 1977, Turner 1999). This may be the reason that the abundance of HC$_3$N  increases as
a function of $\chi$(HNCO). However, as mentioned above, in order to investigate the essential relationship between HNCO and HC$_3$N, high angular resolution observations and chemical models should be carried out in the future.

N$_2$H$^+$ is regarded as an excellent cold gas, as it is more
resistant to freeze-out on grains than carbon-bearing species
(Bergin et al. 2001). Recently, we found that the abundance of
N$_2$H$^+$ has a tight correlation with that of HNC, indicating that
HNC may also be preferentially formed in cold gas (Yu $\&$ Xu 2016).
In a survey of 18 molecular clouds, Jackson et al. (1984) detected
HNCO emissions in seven sources with an average excitation
temperature of 12 K, indicating HNCO could also trace cold gas. The
top middle panel plots the HNC abundance as a function of
$\chi$(HNCO). A least square fit to the data yields $\chi$(HNC) =
0.56 + 0.89 $\chi$(HNCO). The top right panel shows the N$_2$H$^+$
abundance as a function of $\chi$(HNCO), and the functional form of
the linear fit is $\chi$(N$_2$H$^+$) = 0.59 + 0.58 $\chi$(HNCO).

SiO is a well-known shocked gas tracer (e.g. Schilke et al. 1997).
In the ambient gas of energetic young outflows, the abundance of SiO
could jump to almost 10$^{-6}$ (Martin-Pintado et al. 1992). The previous SiO production mechanism explains
spectral lines that show extended wing emission, which is caused by the interaction between high-velocity shocks (typically C-shocks with $\nu >$ 25 km s$^{-1}$ and/or fast J-shocks) and the
surrounding medium. There are also increasing evidences of SiO emission as tracers of low-velocity ($<$ 10 km/s) both in observations and modelling (e.g. Jim\'{e}nez-Serra et al. 2010; Louvet et al. 2016). Jim\'{e}nez-Serra et al. (2010)
detected extended narrow SiO emission (line width of $\sim$ 0.8 km/s) not associated with
signs of star formation in an IRDC. They suggest this narrow line emissions of SiO could be generated by the following processes: i) remnants of large-scale shocks caused by the formation process of the IRDC; ii) decelerated gas in large-scale outflows driven by neighboring massive protostars; iii) undetected and widespread lower mass protostars.  In our ATLASGAL sample, SiO emission is mostly detected in
clumps with signs of star formation (typically associated with 8 $\mu$m sources, 24 $\mu$m sources, and/or extended 4.5 $\mu$m emissions). Outflows have also been detected in some sources (Yu $\&$ Wang 2014). The average SiO line width in the sample is $\sim$ 9 km/s, indicating the presence of outflow activity.
Zinchenko et al. (2000) found the SiO integrated line intensities
correlate well with those of thermal HNCO emission in massive
galactic dense cores, indicating a common production mechanism of
these two species. Rodr\'iguez-Fern\'andez et al. (2010) tested this
hypothesis by observing the L1157 molecular outflow. Their result
indicate shocks actually enhance the HNCO abundance in the
star-forming regions of galactic nuclei. Besides, Li et al. (2013) found possible shock enhancement of HNCO in Orion KL and W75OH.
They regard collisional excitation is likely to be the dominant
excitation mechanism for HNCO emission. We also found the sizes of HNCO clumps are comparable to those of SiO clumps. However, as shown in the
bottom panel of Fig.20, our study indicate that the abundance of HNCO
dose not correlate well with that of SiO. This may be because HNCO
and SiO trace different parts of shocked gas. Flower et al. (1995)
regard gas-phase HNCO mainly enhanced by low-velocity shocks. Using
interferometric data, Blake et al. (1996) found the spatial
distributions of HNCO and SiO emissions are quite different. The average HNCO line width in the sample is $\sim$ 5 km/s, less than that of SiO. Fig. 21 plots the HNCO velocity widths against those of HN$^{13}$C and SiO.
We can see that the velocity width of HNCO tends to be wider than
that of HN$^{13}$C (a tracer of unshocked dense and cold gas) and be narrower
than that of SiO, indicating HNCO traces relatively low-velocity shocks. Fig. 22 shows the
$\chi$(HNCO)/$\chi$(SiO) relative abundance ratios plotted as a
function of the line widths of HNCO (the left panel) and SiO (the
right panel). There is a hint that in the low-velocity shocks traced
by HNCO, the HNCO abundance increases faster than SiO. However, in
the relatively high-velocity shocks traced by SiO, the SiO abundance increases
faster than HNCO. UV radiation induced by high-velocity shocks (Viti
et al. 2002) might destroy HNCO. In our sample, two clumps
(AGAL331.709+00.582 and AGAL331.709+00.602) are known as
``extended green objects" (EGOs) found by Cyganowski et al. (2008).
EGOs are believed to be good candidates of massive young stellar
objects (MYSOs) with outflows. Besides, these two clumps are found
to be embedded in the same IRDC of MSXDC G331.71+00.59 (Yu $\&$ Wang
2013), indicating the same initial chemical conditions. However, we
found that AGAL331.709+00.508 has relatively strong SiO emissions and weak HNCO emissions.
While the situations in AGAL331.709+00.602 are the
opposite (see figure 8 and figure 9). The left panel of Fig. 23
shows the velocity-width (moment 2) map of SiO overlaid with HNCO
integrated intensity contours. It seems that even though these two
clumps locate in the same IRDC, their physical conditions are quite
different: the shocks traced by SiO in AGAL331.709+00.582 are much
faster than that in AGAL331.709+00.602 (see the right panel of Fig. 23). Fast-velocity shocks may be
destroying the molecule of HNCO in AGAL331.709+00.582. We wish high
spacial resolution observations carried out in the future to study
the differences of HNCO and SiO in different part of shocked
gas.

\section{Summary}
Using data from MALT90, we present molecular line study of a sample of ATLASGAL clumps. By comparing
with other molecular species, we found that the abundance of HNCO
correlates well with  HC$_3$N, HNC, C$_2$H,  H$^{13}$CO$^+$ and N$_2$H$^+$. While the correlations between $\chi$(HNCO) and species such as $^{13}$CS, SiO are not so
good. Previous studies indicate HNCO and SiO are good tracers of
interstellar shocks. However, in this study, we found the abundance
of HNCO does not correlate well with that of SiO. We suggest this may be because HNCO traces the low-velocity shocks while SiO traces relatively high-velocity
shocks. We found that in
the low-velocity shocks, the HNCO abundance increases faster than
that of SiO. While in the relatively high-velocity shocks traced by SiO, the SiO
abundance increases faster than that of HNCO. We suggest that in the
infrared dark cloud of MSXDC G331.71+00.59, high-velocity
shocks are destroying the molecule of HNCO.

\section{ACKNOWLEDGEMENTS}
We thank the referee for constructive comments that improve this paper. This paper has made use information from the APEX Telescope Large
Area Survey of the Galaxy. The ATLASGAL project is a collaboration
between the Max-Planck-Gesellschaft, the European Southern
Observatory (ESO) and the Universidad de Chile. This research made
use of data products from the Millimetre Astronomy Legacy Team 90
GHz (MALT90) survey. The Mopra telescope is part of the Australia
Telescope and is funded by the Commonwealth of Australia for
operation as National Facility managed by CSIRO. This paper is
supported by National Natural Science Foundation of China under
grants of 11503037.


\clearpage
\begin{table}
\bc
\begin{minipage}[]{100mm}
\caption[]{List of our sources.}\end{minipage}
\setlength{\tabcolsep}{1pt}
\small
 \begin{tabular}{ccccccccccccc}
  \hline\noalign{\smallskip}
 Source   & RA & Dec    & $T_d$$^a$   &  $N(H_2)$$^a$   & Type$^{a}$  \\
  name    & (J2000.0) & (J2000.0)    & (K)     &($\times$ 10$^{23}$ cm$^{-2}$)\\
  \hline\noalign{\smallskip}
AGAL008.671-00.356 &  18:06:19.13  & -21:37:27.3   &  24.0(3.0)     &  3.54    (0.43) &   Protostellar  \\
AGAL008.684-00.367 &  18:06:23.44  & -21:37:05.9   &  24.6(0.4)     &  1.15    (0.08) &  Protostellar   \\
AGAL010.472+00.027 &  18:08:38.18  & -19:51:49.0   &  30.0(2.0)     &  5.24    (0.51) &   HII           \\
AGAL318.948-00.197 &  15:00:55.54  & -58:58:57.5   &  25.6(0.4)     &  1.41    (0.10) &   Protostellar  \\
AGAL327.293-00.579 &  15:53:08.52  & -54:37:06.9   &  28.0(1.0)     &  8.69    (0.62) &  HII            \\
AGAL329.029-00.206 &  16:00:31.95  & -53:12:53.1   &  19.6(0.7)     &  2.62    (0.19) &   Protostellar  \\
AGAL329.066-00.307 &  16:01:09.75  & -53:16:03.3   &  18.5(0.6)     &  1.07    (0.08) &   Protostellar  \\
AGAL331.709+00.582 &  16:10:01.56  & -50:49:34.8   &  19.0(3.0)     &  1.04    (0.15) &  Protostellar   \\
AGAL331.709+00.602 &  16:10:01.56  & -50:49:34.8   &  20.2(0.9)     &  1.09    (0.08) &   Protostellar  \\
AGAL335.586-00.291 &  16:30:59.08  & -48:43:53.3   &  22.5(0.4)     &  2.75    (0.20) &   Protostellar  \\
AGAL337.704-00.054 &  16:38:29.64  & -47:00:41.1   &  22.6(0.9)     &  3.08    (0.22) &   HII           \\
AGAL338.926+00.554 &  16:40:34.29  & -45:41:41.8   &  20.0(1.0)     &  4.78    (0.34) &   Protostellar  \\
AGAL340.248-00.374 &  16:49:30.41  & -45:17:53.6   &  22.0(1.0)     &  1.17    (0.08) &   HII           \\
AGAL345.003-00.224 &  17:05:11.02  & -41:29:07.8   &  25.8(11.0)    &  3.38    (0.88) &   Protostellar  \\
AGAL350.111+00.089 &  17:19:26.61  & -37:10:23.1   &  24.0(0.6)     &  1.55    (0.07) &  HII            \\
AGAL351.444+00.659 &  17:20:54.64  & -35:45:11.8   &  22.0(2.0)     &  9.75    (0.94) &  Protostellar   \\
AGAL351.581-00.352 &  17:25:24.99  & -36:12:45.1   &  24.0(1.0)     &  5.61    (0.40) &   Protostellar  \\
AGAL353.409-00.361 &  17:30:26.21  & -34:41:48.9   &  23.0(1.0)     &  4.89    (0.35) &   HII           \\
  \noalign{\smallskip}\hline
\end{tabular}
\ec
\tablecomments{0.86\textwidth}{a: These values come from Guzm\'{a}n et al. (2015).}
\end{table}

\begin{table}
\bc
\begin{minipage}[]{100mm}
\caption[]{The beam deconvolved angular diameters of molecular clouds.}\end{minipage}
\setlength{\tabcolsep}{1pt}
\small
 \begin{tabular}{ccccccccccccc}
  \hline\noalign{\smallskip}
Source   & $\theta$(HNCO)  &  $\theta$(HC$_3$N) &  $\theta$(N$_2$H$^+$) &   $\theta$(HN$^{13}$C)   &  $\theta$($^{13}$CS)  &  $\theta$(C$_2$H) & $\theta$(H$^{13}$CO$^+$)  & $\theta$(SiO)\\
name   & arcsec       & arcsec     & arcsec      & arcsec      &     arcsec         & arcsec      & arcsec & arcsec \\
  \hline\noalign{\smallskip}
AGAL008.671-00.356      &   71.38       &   ...      & 79.34    & ...       & ...    &   ...       & ...         & 66.45   \\
AGAL008.684-00.367      &   ...         &   ...      & ...      & ...       & ...    &   ...       & ...         & ...      \\
AGAL010.472+00.027      &   60.26       &   16.18    & 58.91    & 37.29     & ...    &   40.13     & 32.72       & 46.62     \\
AGAL318.948-00.197      &   36.00       &   17.93    & 40.50    & 33.96     & ...    &   47.65     & 43.41       & 23.90     \\
AGAL327.293-00.579      &   19.66       &   50.85    & 54.79    & 63.99     & 30.46  &   118.96    & 72.04       & 33.96        \\
AGAL329.029-00.206      &   47.58       &   38.55    & 46.13    & 47.21     & ...    &   51.43     & 27.26       & 52.69     \\
AGAL329.066-00.307      &   37.29       &   45.28    & 52.12    & 39.46     & ...    &   63.39     & 25.35       & 53.07     \\
AGAL331.709+00.582      &   ...         &   ...      & ...      & ...       & ...    &   ...       & ...         & ...          \\
AGAL331.709+00.602      &   30.55       &   28.85    & ...      & ...       & ...    &   28.74     & ...         & 12.44     \\
AGAL335.586-00.291      &   39.99       &   30.36    & 39.54    & 8.78      & ...    &   33.34     & 34.74       & 33.43     \\
AGAL337.704-00.054      &   28.11       &   11.97    & 21.40    & 34.13     & ...    &   38.70     & 34.74       & 24.74     \\
AGAL338.926+00.554      &   28.02       &   41.01    & 51.20    & 34.83     & ...    &   30.94     & 64.78       & 41.66     \\
AGAL340.248-00.374      &   42.22       &   47.40    & 60.95    & 25.58     & ...    &   70.15     & 66.42       & 40.65     \\
AGAL345.003-00.224      &   37.13       &   32.25    & 89.05    & 44.67     & 6.91   &   33.53     & 30.84       & 42.09     \\
AGAL350.111+00.089      &   55.86       &   ...      & ...      & ...       & ...    &   ...       & ...         & 45.42        \\
AGAL351.444+00.659      &   46.90       &   42.58    & 68.95    & 44.63     & 67.12  &   103.89    & 49.91       & 34.58        \\
AGAL351.581-00.352      &   67.40       &   29.16    & 81.56    & ...       & 52.00  &   40.21     & 53.73       & 39.15     \\
AGAL353.409-00.361      &   45.03       &   39.77    & 88.64    & 68.92     & 42.30  &   80.64     & 64.96       & 39.32     \\
  \noalign{\smallskip}\hline
\end{tabular}
\ec
\end{table}

\begin{table}
\bc
\begin{minipage}[]{100mm}
\caption[]{ Line widths of species.}\end{minipage}
\setlength{\tabcolsep}{1pt}
\small
 \begin{tabular}{ccccccccccccc}
  \hline\noalign{\smallskip}
Source   & HNCO       &  HC$_3$N   &  N$_2$H$^+$ &    HN$^{13}$C      &     $^{13}$CS      &    C$_2$H   & H$^{13}$CO$^+$  & SiO\\
name   & km s$^{-1}$  & km s$^{-1}$  & km s$^{-1}$  & km s$^{-1}$   & km s$^{-1}$         & km s$^{-1}$  & km s$^{-1}$ & km s$^{-1}$ \\
  \hline\noalign{\smallskip}
AGAL008.671-00.356      &   6.24    (0.54)       &   4.48   (0.14  )   & 3.47   (0.09  )   & 4.33   (0.31  )    & 5.20  (0.77 ) &    5.25   (0.35  )   & 2.38   (0.24  )   & ...               \\
AGAL008.684-00.367      &   6.66    (0.69)       &    3.65  (0.16  )   & 5.31   (0.09  )   & 4.66   (0.47  )    & ...           &    4.59   (0.32  )   & 4.21   (0.28 )      & ...                \\
AGAL010.472+00.027      &   6.52    (0.65)       &   7.24   (0.26  )   & 4.76   (0.37  )   & 6.74   (0.64  )    & 7.87  (0.78 ) &    7.75   (0.45  )   & 6.37   (0.42  )   &  6.34 (0.72    )    \\
AGAL318.948-00.197      &   4.56    (0.79)       &   3.07   (0.17  )   & 3.06   (0.05  )   & 3.02   (0.49  )    & 3.06  (0.64 ) &    2.76   (0.19  )   & 2.63   (0.17  )   &  8.55 (0.81    )    \\
AGAL327.293-00.579      &   5.45    (0.65)       &    5.89  (0.10  )   & 5.56   (0.18  )   & 5.14   (0.30  )    & 6.18  (0.21 ) &    6.20   (0.20  )   & 4.99   (0.25 )      & 7.54 (0.46   )       \\
AGAL329.029-00.206      &   3.85    (0.36)       &   4.90   (0.17  )   & 5.79   (0.17  )   & 3.37   (0.34  )    & ...           &    5.13   (0.54  )   & 4.50   (0.45  )   &  10.31(    0.65  )    \\
AGAL329.066-00.307      &   4.41    (0.51)       &   5.39   (0.43  )   & 5.42   (0.19  )   & 5.51   (0.60  )    & ...           &    4.27   (0.50  )   & 5.98   (0.64  )   &  9.79 (0.97    )    \\
AGAL331.709+00.582      &   2.76    (0.49)       &    4.83  (0.33  )   & 5.01   (0.14  )   & 3.71   (0.45  )    & ...           &    4.37   (0.55  )   & 4.06   (0.51 )      & 12.07(   1.33  )       \\
AGAL331.709+00.602      &   4.42    (0.57)       &   3.45   (0.18  )   & 3.94   (0.09  )   & 2.60   (0.30  )    & ...           &    4.08   (0.44  )   & 3.39   (0.38  )   &  6.02 (0.76    )    \\
AGAL335.586-00.291      &   3.52    (0.33)       &   3.54   (0.12  )   & 3.27   (0.05  )   & 3.41   (0.29  )    & ...           &    3.40   (0.22  )   & 3.55   (0.15  )   &  12.43(    0.90  )    \\
AGAL337.704-00.054      &   7.19    (0.69)       &   6.72   (0.28  )   & 6.21   (0.43  )   & 5.98   (0.65  )    & 5.75  (0.74 ) &    6.70   (0.39  )   & 4.87   (0.40  )   &  8.96 (0.80    )    \\
AGAL338.926+00.554      &   6.46    (0.53)       &   6.42   (0.31  )   & 7.78   (0.36  )   & 7.02   (0.56  )    & ...           &    6.83   (0.46  )   & 7.30   (0.46  )   &  8.28 (0.66    )    \\
AGAL340.248-00.374      &   3.97    (0.40)       &   3.28   (0.14  )   & 3.78   (0.08  )   & 2.95   (0.22  )    & ...           &    3.64   (0.29  )   & 2.94   (0.18  )   &  8.27 (0.89    )    \\
AGAL345.003-00.224      &   10.96   (1.14)       &   6.95   (0.20  )   & 4.10   (0.14  )   & 5.26   (0.39  )    & 7.39  (1.03 ) &    6.14   (0.47  )   & 5.58   (0.46  )   &  16.78(    0.78  )    \\
AGAL350.111+00.089      &   7.33    (0.66)       &    5.52  (0.39  )   & 4.67   (0.10  )   & 5.11   (0.56  )    & 2.29  (0.56 ) &    7.84   (0.45  )   & 6.78   (0.53 )      & 9.53 (1.15   )       \\
AGAL351.444+00.659      &   4.69    (0.31)       &    3.98  (0.06  )   & 4.65   (0.43  )   & 3.78   (0.11  )    & 4.05  (0.16 ) &    3.94   (0.05  )   & 3.99   (0.08 )      & 8.35 (0.21   )       \\
AGAL351.581-00.352      &   4.9     (0.33)       &   5.15   (0.15  )   & 3.13   (0.20  )   & 5.80   (1.05  )    & 5.11  (0.67 ) &    7.30   (0.64  )   & 5.36   (0.59  )   &  8.35 (1.01    )    \\
AGAL353.409-00.361      &   5.95    (0.42)       &   4.81   (0.13  )   & 5.00   (0.43  )   & 4.72   (0.27  )    & 4.61  (0.43 ) &    6.47   (0.24  )   & 4.46   (0.17  )   &  7.15 (0.41    )    \\
  \noalign{\smallskip}\hline
\end{tabular}
\ec
\end{table}

\clearpage

\begin{table}
\bc
\begin{minipage}[]{100mm}
\caption[]{ Peak intensities of species.}\end{minipage}
\setlength{\tabcolsep}{1pt}
\small
 \begin{tabular}{ccccccccccccc}
  \hline\noalign{\smallskip}
Source   & HNCO       &  HC$_3$N   &  N$_2$H$^+$ &    HN$^{13}$C      &     $^{13}$CS      &    C$_2$H   & H$^{13}$CO$^+$  & SiO\\
name   & K          & K  & K  & K  & K       & K & K & K \\
  \hline\noalign{\smallskip}
AGAL008.671-00.356      &   0.78       &   2.25    & 4.51    & 1.04     & 0.52 &    1.40    & 1.33       & ...         \\
AGAL008.684-00.367      &   0.65       &    2.04   & 3.46    & 0.85     &...   &    1.08    & 1.01       & ...         \\
AGAL010.472+00.027      &   0.55       &   1.37    & 2.19    & 0.54     & 0.51 &    0.99    & 0.77       &  0.52       \\
AGAL318.948-00.197      &   0.41       &   1.62    & 5.06    & 0.67     & 0.35 &    1.39    & 1.42       &  0.62       \\
AGAL327.293-00.579      &   0.61       &    3.52   & 5.38    & 0.93     & 1.52 &    2.29    & 1.38       & 1.14        \\
AGAL329.029-00.206      &   0.88       &   2.09    & 3.33    & 0.84     &...   &    0.76    & 0.67       &  0.97       \\
AGAL329.066-00.307      &   0.50       &   0.89    & 2.75    & 0.59     &...   &    0.61    & 0.56       &  0.52       \\
AGAL331.709+00.582      &   0.55       &    0.97   & 2.86    & 0.52     &...   &    0.55    & 0.50       & 0.50        \\
AGAL331.709+00.602      &   0.85       &   1.32    & 3.81    & 0.95     &...   &    0.82    & 0.84       &  0.63       \\
AGAL335.586-00.291      &   0.87       &   2.18    & 6.10    & 1.08     &...   &    1.30    & 1.78       &  0.76       \\
AGAL337.704-00.054      &   0.55       &   1.28    & 3.13    & 0.44     & 0.53 &    1.07    & 0.72       &  0.64       \\
AGAL338.926+00.554      &   0.79       &   1.41    & 5.82    & 0.71     &...   &    0.94    & 1.08       &  0.73       \\
AGAL340.248-00.374      &   0.70       &   1.81    & 6.60    & 1.07     &...   &    1.18    & 1.43       &  0.53       \\
AGAL345.003-00.224      &   0.34       &   1.67    & 2.91    & 0.68     & 0.39 &    0.74    & 0.76       &  0.93       \\
AGAL350.111+00.089      &   0.64       &    1.51   & 5.25    & 0.69     & 0.51 &    1.22    & 0.83       & 0.59        \\
AGAL351.444+00.659      &   1.27       &    7.54   & 15.06   & 2.60     & 2.14 &    6.85    & 3.47       & 2.43        \\
AGAL351.581-00.352      &   1.25       &   2.11    & 1.94    & 0.48     & 0.59 &    0.68    & 0.76       &  0.46       \\
AGAL353.409-00.361      &   0.73       &   2.73    & 6.03    & 1.39     & 0.73 &    2.25    & 1.97       &  1.31       \\
  \noalign{\smallskip}\hline
\end{tabular}
\ec
\end{table}

\clearpage

\begin{table}
\bc
\begin{minipage}[]{100mm}
\caption[]{Integrated intensities of species.}\end{minipage}
\setlength{\tabcolsep}{1pt}
\small
 \begin{tabular}{ccccccccccccc}
  \hline\noalign{\smallskip}
Source   & HNCO       &  HC$_3$N   &  N$_2$H$^+$ &    HN$^{13}$C      &     $^{13}$CS      &    C$_2$H   & H$^{13}$CO$^+$  & SiO\\
name   & K km s$^{-1}$  & K km s$^{-1}$  & K km s$^{-1}$  & K km s$^{-1}$  & K km s$^{-1}$         & K km s$^{-1}$  & K km s$^{-1}$ & K km s$^{-1}$ \\
  \hline\noalign{\smallskip}
AGAL008.671-00.356      &   5.22    (0.37  )      &   10.77 (   0.29  )    & 16.67  (0.69  )   & 4.81   (0.29  )    & 2.91  (0.32  )   &    7.87    (0.41    )   & 3.39     (0.69  )      &...                \\
AGAL008.684-00.367      &   4.64    (0.40 )       &    7.95     (0.27  )   & 19.59  (0.02  )   & 4.24   (0.34  )    & ...              &    5.32    (0.33    )   & 4.56     (0.28  )      &...                 \\
AGAL010.472+00.027      &   3.88    (0.33  )      &   10.58 (   0.32  )    & 11.13  (1.58  )   & 3.93   (0.32  )    & 4.30  (0.36  )   &    8.17    (0.39    )   & 5.26     (0.31  )      &  3.54   (0.34  )    \\
AGAL318.948-00.197      &   2.02    (0.29  )      &   5.30  (0.25   )      & 16.52    (0.25)   & 2.16   (0.26  )    & 1.49  (0.24  )   &    4.09    (0.23    )   & 3.98     (0.22  )      &  5.67   (0.42  )    \\
AGAL327.293-00.579      &   3.57    (0.33 )       &    22.12    (0.33  )   & 31.95  (1.66  )   & 5.14   (0.28  )    & 10.01 (   0.31 ) &    15.13 (   0.44   )   & 7.34     (0.33  )      & 9.22 (0.46    )       \\
AGAL329.029-00.206      &   3.63    (0.29  )      &   10.91 (   0.31  )    & 20.55  (0.53  )   & 3.03   (0.25  )    & ...              &    4.17    (0.35    )   & 3.22     (0.30  )      &  10.73  (0.51  )    \\
AGAL329.066-00.307      &   2.39    (0.23  )      &   4.96  (0.33     )    & 15.91    (0.55)   & 3.49   (0.30  )    & ...              &    2.78    (0.25    )   & 3.57     (0.32  )      &  5.42   (0.42  )    \\
AGAL331.709+00.582      &   1.64    (0.21 )       &    4.62     (0.27  )   & 15.25  (0.37  )   & 2.08   (0.23  )    & ...              &    2.57    (0.27    )   & 2.19     (0.25  )      & 6.50 (0.55    )       \\
AGAL331.709+00.602      &   4.03    (0.38  )      &   4.87  (0.23    )     & 16.00  (0.37  )   & 2.65   (0.25  )    & ...              &    3.59    (0.31    )   & 3.04     (0.29  )      &  4.05   (0.38  )    \\
AGAL335.586-00.291      &   3.26    (0.25  )      &   8.22  (0.23   )      & 21.24  (0.30  )   & 3.94   (0.29  )    & ...              &    4.72    (0.26    )   & 6.75     (0.25  )      &  10.13  (0.53  )    \\
AGAL337.704-00.054      &   4.21    (0.34  )      &   9.20  (0.33   )      & 20.73  (0.50  )   & 2.83   (0.27  )    & 3.29  (0.32  )   &    7.63    (0.36    )   & 3.78     (0.28  )      &  6.15   (0.42  )    \\
AGAL338.926+00.554      &   5.48    (0.41  )      &   9.65  (0.39   )      & 48.24  (2.99  )   & 5.33   (0.41  )    & ...              &    6.85    (0.42    )   & 8.44     (0.45  )      &  6.45   (0.45  )    \\
AGAL340.248-00.374      &   2.97    (0.24  )      &   6.33  (0.22   )      & 26.60  (0.46  )   & 3.38   (0.23  )    & ...              &    4.57    (0.28    )   & 4.48     (0.23  )      &  4.75   (0.40  )    \\
AGAL345.003-00.224      &   4.08    (0.35  )      &   12.42 (   0.29  )    & 12.74  (0.53  )   & 3.82   (0.24  )    & 3.09  (0.35  )   &    4.88    (0.32    )   & 4.56     (0.31  )      &  16.66  (0.62  )    \\
AGAL350.111+00.089      &   5.06    (0.39 )       &    8.88     (0.46  )   & 26.17  (0.54  )   & 3.78   (0.31  )    & 1.26  (0.24    ) &    10.19 ( 0.44   )     & 6.04     (0.39  )      & 6.01 (0.50    )       \\
AGAL351.444+00.659      &   6.34    (0.34 )       &    32.02    (0.45  )   & 74.65  (2.36  )   & 10.49  (0.27  )    & 9.20  (0.30    ) &    28.78 ( 0.31   )     & 14.79    (0.26  )      & 21.64(  0.44  )       \\
AGAL351.581-00.352      &   6.53    (0.34  )      &   11.58 (   0.30  )    & 6.49   (0.50  )   & 2.98   (0.37  )    & 3.21  (0.35  )   &    5.30    (0.42    )   & 4.35     (0.37  )      &  4.11   (0.43  )    \\
AGAL353.409-00.361      &   4.63    (0.37  )      &   14.02 (   0.31  )    & 32.11  (1.14  )   & 6.99   (0.32  )    & 3.59  (0.28  )   &    15.51 ( 0.49   )     & 9.40     (0.30  )      &  9.99   (0.41  )    \\
  \noalign{\smallskip}\hline
\end{tabular}
\ec
\end{table}

\clearpage

\begin{table}
\bc
\begin{minipage}[]{100mm}
\caption[]{Column densities of species.}\end{minipage}
\setlength{\tabcolsep}{1pt}
\small
 \begin{tabular}{ccccccccccccc}
  \hline\noalign{\smallskip}
Source   & HNCO               &      HC$_3$N     &      N$_2$H$^+$        &       HNC             &     $^{13}$CS               &    C$_2$H           & H$^{13}$CO$^+$       & SiO\\
name   & 10$^{13}$ cm$^{-2}$  & 10$^{13}$ cm$^{-2}$  & 10$^{13}$ cm$^{-2}$ & 10$^{14}$ cm$^{-2}$  & 10$^{13}$ cm$^{-2}$         & 10$^{15}$ cm$^{-2}$ & 10$^{13}$ cm$^{-2}$ & 10$^{13}$ cm$^{-2}$\\
  \hline\noalign{\smallskip}
AGAL008.671-00.356      &   9.63    (2.01)       &   4.33   (0.11)   & 9.75     (1.33)   & 7.69     (0.94)    & 2.13    (0.44) &   1.40     (0.21)   &  6.19    (1.82)      &...                \\
AGAL008.684-00.367      &   8.77    (0.15)       &    3.19  (0.11)   & 5.55     (0.07)   & 9.66     (0.47)    & ...            &    1.12    (0.11)   & 8.51     (0.63)      &...                 \\
AGAL010.472+00.027      &   9.05    (2.00)       &   4.29   (0.18)   & 7.66     (1.65)   & 7.30     (0.39)    & 3.70    (0.51) &   1.00     (0.10)   &  10.57   (1.19)      &  1.81     (0.27)    \\
AGAL318.948-00.197      &   3.98    (0.64)       &   2.12   (0.10)   & 3.43     (0.10)   & 5.28     (1.27)    & 1.14    (0.20) &   0.44     (0.03)   &  7.82    (0.53)      &  2.59     (0.23)    \\
AGAL327.293-00.579      &   7.73    (1.04)       &    8.91  (0.17)   & 8.53     (0.69)   & 10.94    (0.37)    & 8.19    (0.48) &    2.19    (0.12)   & 16.22    (1.18)      & 4.49  (0.35)       \\
AGAL329.029-00.206      &   5.53    (0.65)       &   4.55   (0.08)   & 6.78     (0.34)   & 11.72    (0.88)    & ...            &   0.63     (0.07)   &  5.69    (0.67)      &  4.14     (0.29)    \\
AGAL329.066-00.307      &   3.46    (0.44)       &   2.10   (0.11)   & 2.51     (0.14)   & 3.48     (0.14)    & ...            &   0.67     (0.07)   &  5.60    (0.63)      &  2.02     (0.20)    \\
AGAL331.709+00.582      &   2.43    (0.7 )       &    1.92  (0.03)   & 2.32     (0.30)   & 3.78     (0.81)    & ...            &    0.51    (0.11)   & 3.31     (0.75)      & 2.46  (0.46)       \\
AGAL331.709+00.602      &   6.30  (0.88)         &   2.01   (0.07)   & 3.59     (0.19)   & 5.41     (0.28)    & ...            &   0.53     (0.07)   &  5.06    (0.64)      &  1.59     (0.19)    \\
AGAL335.586-00.291      &   5.64    (0.54)       &   3.32   (0.09)   & 5.98     (0.16)   & 5.12     (0.31)    & ...            &   0.46     (0.03)   &  10.91   (0.55)      &  4.26     (0.27)    \\
AGAL337.704-00.054      &   7.32    (0.9 )       &   3.72   (0.11)   & 5.34     (0.28)   & 4.02     (0.14)    & 2.31    (0.30) &   1.46     (0.11)   &  7.10    (0.74)      &  2.59     (0.25)    \\
AGAL338.926+00.554      &   8.50  (1.05)         &   4.00   (0.11)   & 7.60     (0.74)   & 6.02     (0.37)    & ...            &   1.03     (0.10)   &  12.88   (1.12)      &  2.52     (0.25)    \\
AGAL340.248-00.374      &   5.03    (0.64)       &   2.57   (0.07)   & 5.53     (0.27)   & 7.77     (0.60)    & ...            &   0.58     (0.08)   &  8.27    (0.69)      &  1.96     (0.23)    \\
AGAL345.003-00.224      &   8.10  (4.95)         &   4.98   (0.43)   & 4.51     (1.71)   & 12.60    (4.19)    & 2.38    (1.09) &   0.58     (0.24)   &  9.93    (4.00)      &  7.67     (2.68)    \\
AGAL350.111+00.089      &   9.33    (0.7 )       &    3.57  (0.18)   & 6.57     (0.25)   & 8.93     (0.22)    & 0.92    (0.19) &    1.05    (0.06)   & 10.92    (0.91)      & 2.63  (0.27)       \\
AGAL351.444+00.659      &   10.74   (1.58)       &    13.01 (0.05)   & 24.60    (2.40)   & 22.23    (1.18)    & 6.37    (0.60) &    3.61    (0.27)   & 26.43    (2.09)      & 8.97  (0.71)       \\
AGAL351.581-00.352      &   12.05   (1.16)       &   4.65   (0.11)   & 5.72     (0.62)   & 3.93     (0.34)    & 2.35    (0.33) &   1.86     (0.21)   &  8.52    (1.00)      &  1.80     (0.24)    \\
AGAL353.409-00.361      &   8.18    (1.03)       &   5.66   (0.10)   & 10.07    (0.68)   & 14.06    (0.76)    & 2.56    (0.28) &   2.28     (0.15)   &  16.43   (1.03)      &  4.26     (0.30)    \\
  \noalign{\smallskip}\hline
\end{tabular}
\ec
\end{table}

\clearpage

\begin{table}
\bc
\begin{minipage}[]{100mm}
\caption[]{Abundances of species.}\end{minipage}
\setlength{\tabcolsep}{1pt}
\small
 \begin{tabular}{ccccccccccccc}
  \hline\noalign{\smallskip}
Source   & HNCO       &  HC$_3$N   &  N$_2$H$^+$ &    HNC      &     $^{13}$CS      &    C$_2$H   & H$^{13}$CO$^+$  & SiO\\
name   & 10$^{-10}$  & 10$^{-10}$  & 10$^{-10}$  & 10$^{-9}$   & 10$^{-10}$         & 10$^{-8}$  & 10$^{-10}$ & 10$^{-10}$ \\
  \hline\noalign{\smallskip}
AGAL008.671-00.356      &   2.72    (1.02 )      &   1.22   (0.21 )   & 2.75    (0.81 )   & 2.17    (0.60 )    & 0.60     (0.23 ) &    0.40     (0.12 )   & 1.75    (0.83 )      & ...               \\
AGAL008.684-00.367      &   7.65    (0.73)       &    2.78  (0.32 )   & 4.84    (0.44 )   & 8.43    (1.09 )    & ...              &    0.98     (0.20 )   & 7.43    (1.16 )      & ...                \\
AGAL010.472+00.027      &   1.73    (0.61 )      &   0.82   (0.13 )   & 1.46    (0.50 )   & 1.39    (0.23 )    & 0.71     (0.18 ) &    0.19     (0.04 )   & 2.02    (0.47 )      &  0.35    (0.09 )    \\
AGAL318.948-00.197      &   2.82    (0.71 )      &   1.50   (0.19 )   & 2.43    (0.26 )   & 3.75    (1.26 )    & 0.81     (0.22 ) &    0.31     (0.05 )   & 5.55    (0.83 )      &  1.84    (0.32 )    \\
AGAL327.293-00.579      &   0.89    (0.20)       &    1.03  (0.10 )   & 0.98    (0.16 )   & 1.26    (0.14 )    & 0.94     (0.13 ) &    0.25     (0.03 )   & 1.87    (0.29 )      & 0.52     (0.08 )       \\
AGAL329.029-00.206      &   2.11    (0.43 )      &   1.73   (0.17 )   & 2.58    (0.34 )   & 4.37    (0.79 )    & ...              &    0.24     (0.05 )   & 2.17    (0.44 )      &  1.58    (0.24 )    \\
AGAL329.066-00.307      &   3.24    (0.69 )      &   1.96   (0.28 )   & 2.35    (0.32 )   & 3.25    (0.39 )    & ...              &    0.63     (0.12 )   & 5.24    (1.04 )      &  1.89    (0.35 )    \\
AGAL331.709+00.582      &   2.33    (1.19)       &    1.80  (0.39 )   & 2.22    (0.72 )   & 3.62    (1.54 )    & ...              &    0.49     (0.21 )   & 3.17    (1.39 )      & 2.35     (0.93 )       \\
AGAL331.709+00.602      &   5.76    (1.31)       &   1.84   (0.21 )   & 3.28    (0.44 )   & 4.94    (0.66 )    & ...              &    0.48     (0.11 )   & 4.62    (0.99 )      &  1.45    (0.30 )    \\
AGAL335.586-00.291      &   2.05    (0.37 )      &   1.21   (0.13 )   & 2.18    (0.23 )   & 1.86    (0.26 )    & ...              &    0.17     (0.02 )   & 3.97    (0.52 )      &  1.55    (0.23 )    \\
AGAL337.704-00.054      &   2.37    (0.50 )      &   1.21   (0.13 )   & 1.73    (0.23 )   & 1.30    (0.15 )    & 0.75     (0.16 ) &    0.47     (0.07 )   & 2.30    (0.44 )      &  0.84    (0.15 )    \\
AGAL338.926+00.554      &   1.78    (0.37 )      &   0.84   (0.09 )   & 1.59    (0.29 )   & 1.26    (0.18 )    & ...              &    0.22     (0.04 )   & 2.70    (0.46 )      &  0.53    (0.10 )    \\
AGAL340.248-00.374      &   4.29    (0.92 )      &   2.19   (0.23 )   & 4.72    (0.61 )   & 6.63    (1.06 )    & ...              &    0.49     (0.11 )   & 7.05    (1.18 )      &  1.67    (0.34 )    \\
AGAL345.003-00.224      &   2.40    (2.81 )      &   1.47   (0.69 )   & 1.33    (1.15 )   & 3.73    (2.97 )    & 0.70     (0.68 ) &    0.17     (0.16 )   & 2.94    (2.62 )      &  2.27    (1.86 )    \\
AGAL350.111+00.089      &   6.04    (0.77)       &    2.31  (0.24 )   & 4.25    (0.38 )   & 5.77    (0.37 )    & 0.60     (0.16 ) &    0.68     (0.07 )   & 7.07    (0.97 )      & 1.70     (0.27 )       \\
AGAL351.444+00.659      &   1.10    (0.30)       &    1.33  (0.15 )   & 2.52    (0.54 )   & 2.28    (0.38 )    & 0.65     (0.14 ) &    0.37     (0.07 )   & 2.71    (0.53 )      & 0.92     (0.18 )       \\
AGAL351.581-00.352      &   2.15    (0.39 )      &   0.83   (0.08 )   & 1.02    (0.20 )   & 0.70    (0.12 )    & 0.42     (0.10 ) &    0.33     (0.07 )   & 1.52    (0.31 )      &  0.32    (0.07 )    \\
AGAL353.409-00.361      &   1.67    (0.27 )      &   1.16   (0.11 )   & 2.06    (0.31 )   & 2.88    (0.39 )    & 0.52     (0.10 ) &    0.47     (0.07 )   & 3.36    (0.49 )      &  0.87    (0.13 )    \\
  \noalign{\smallskip}\hline
\end{tabular}
\ec
\end{table}

\begin{table}
\bc
\begin{minipage}[]{100mm}
\caption[]{Linear fitting results for the correlation between HNCO and other species.}\end{minipage}
\setlength{\tabcolsep}{1pt}
\small
 \begin{tabular}{ccccccc}
  \hline\noalign{\smallskip}
No.  &   linear fitting functions &  Pearson correlation coefficient   \\
  \noalign{\smallskip}\hline
1 & $\chi$(HC$_3$N) = 0.89 + 0.24 $\chi$(HNCO)    & 0.85 \\
2 & $\chi$(HNC) = 0.56 + 0.89 $\chi$(HNCO)        & 0.86 \\
3 & $\chi$(N$_2$H$^+$) = 0.59 + 0.58 $\chi$(HNCO) & 0.87 \\
4 & $\chi$($^{13}$CS) = 0.89 - 0.04 $\chi$(HNCO) & -0.25\\
5 & $\chi$(C$_2$H) = 0.19 + 0.09 $\chi$(HNCO)     & 0.81 \\
6 & $\chi$(H$^{13}$CO$^+$) = 1.23 + 0.88 $\chi$(HNCO) & 0.80\\
7 & $\chi$(SiO) = 0.38 + 0.22 $\chi$(HNCO)        & 0.49 \\

  \noalign{\smallskip}\hline
\end{tabular}
\ec
\end{table}

\clearpage
\begin{figure}
\psfig{file=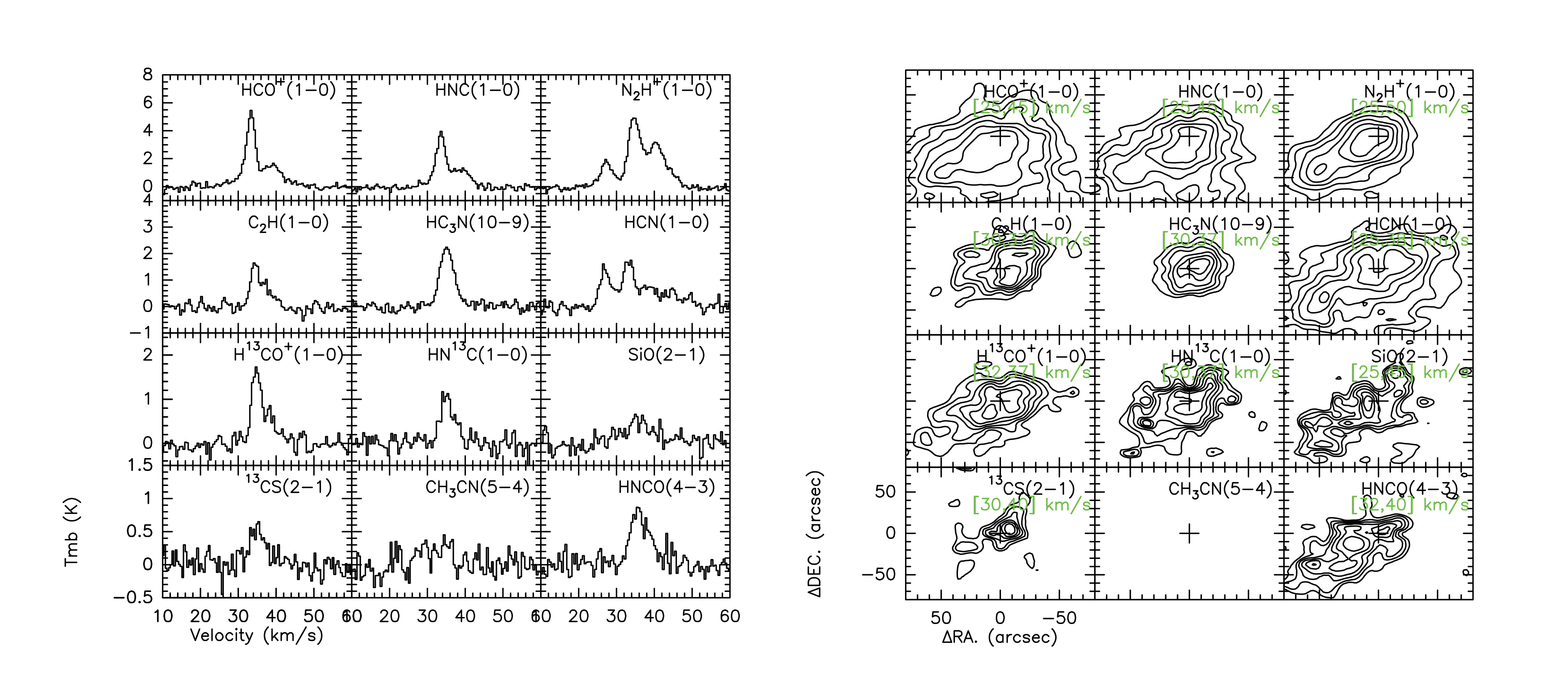,width=6in,height=3in} \caption{Molecular spectra
and integrated intensities of AGAL08.671-00.356. Contour levels are
40, 50, . . . , 90 percent of the peak emissions. The black plus
marks the 870 $\mu$m emission peak of AGAL08.671-00.356. The green number show the range of integrations.}
\end{figure}

\begin{figure}
\psfig{file=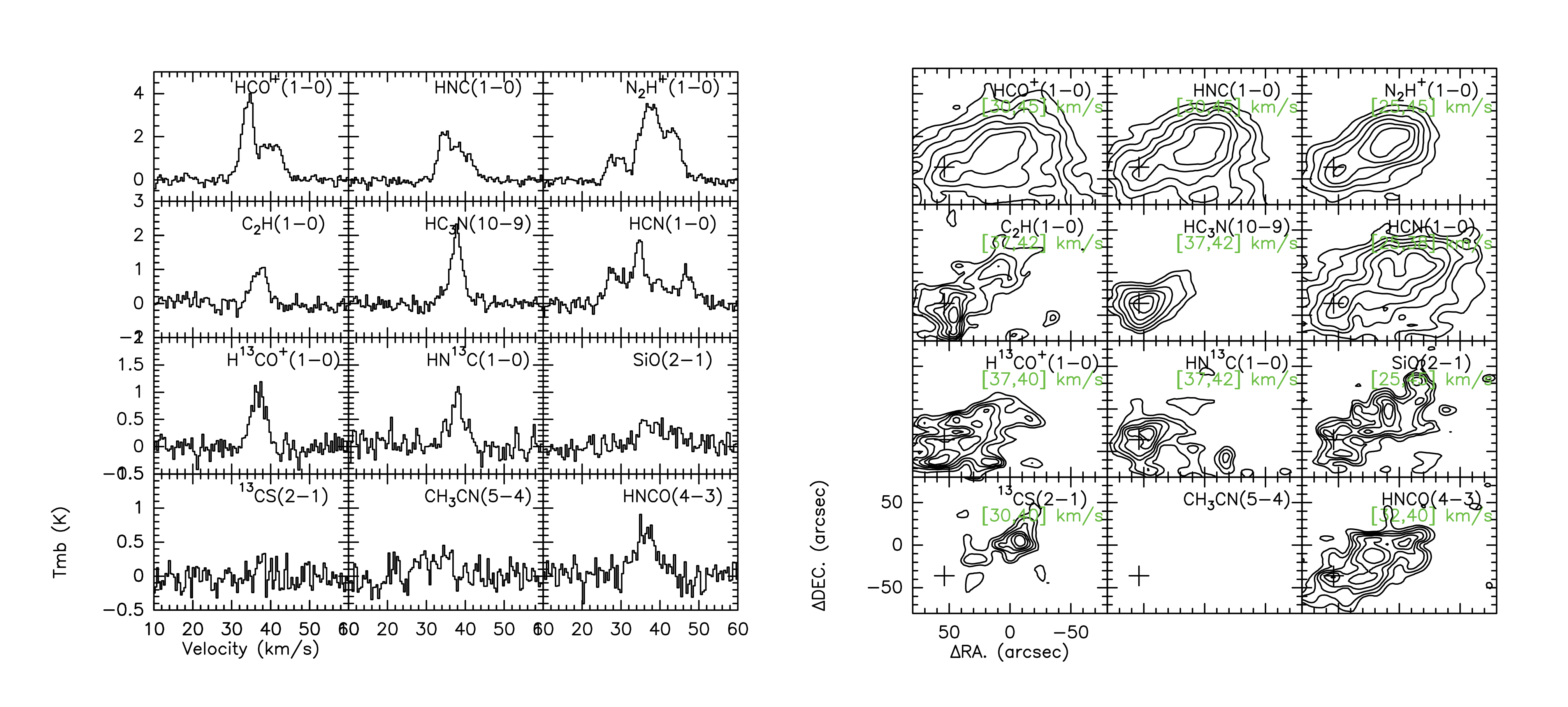,width=6in,height=3in}
 \caption{Molecular spectra
and integrated intensities of AGAL08.684-00.367. Contour levels are
40, 50, . . . , 90 percent of the peak emissions. The black plus
marks the 870 $\mu$m emission peak of AGAL08.684-00.367. The green number show the range of integrations.}
\end{figure}

\begin{figure}
\psfig{file=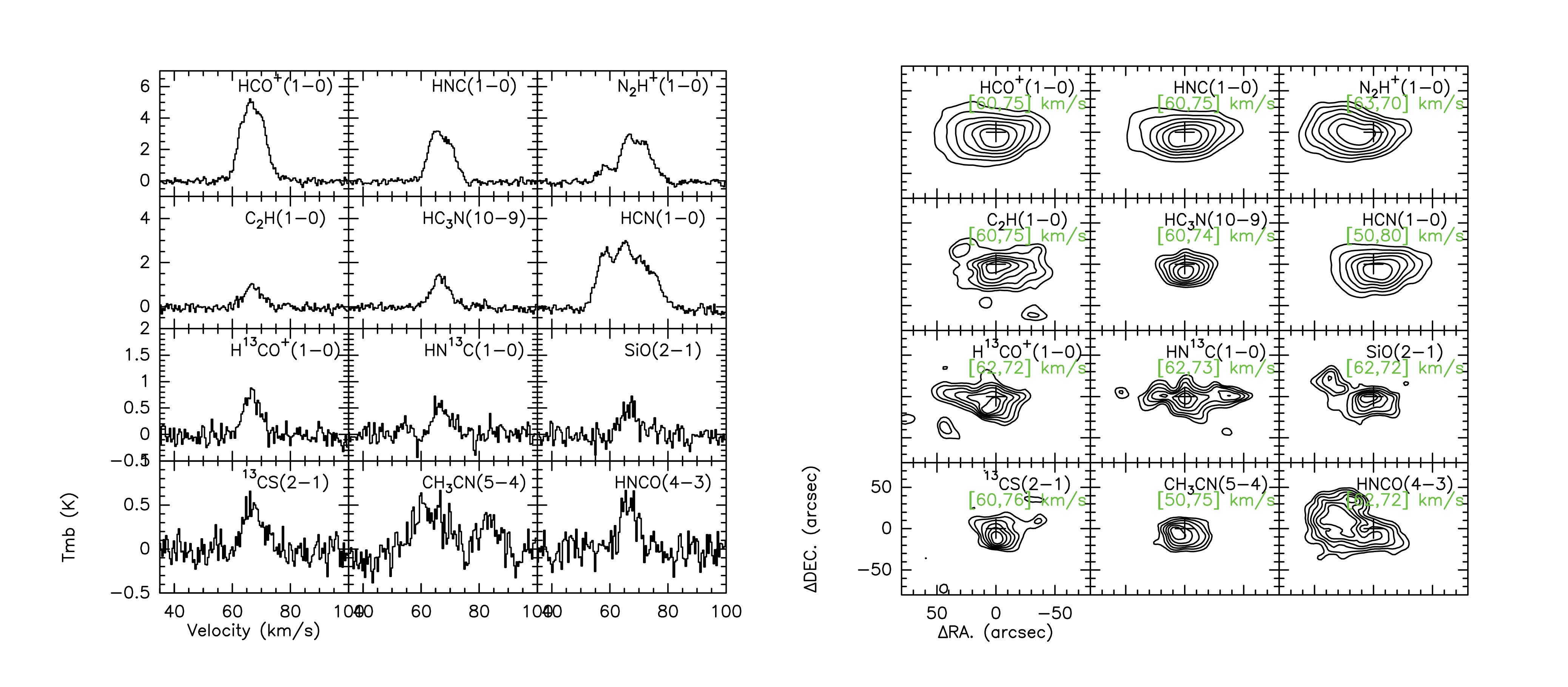,width=6in,height=3in}
 \caption{Molecular spectra
and integrated intensities of AGAL10.472+00.027. Contour levels are
40, 50, . . . , 90 percent of the peak emissions. The black plus
marks the 870 $\mu$m emission peak of AGAL10.472+00.027. The green number show the range of integrations.}
\end{figure}

\begin{figure}
\psfig{file=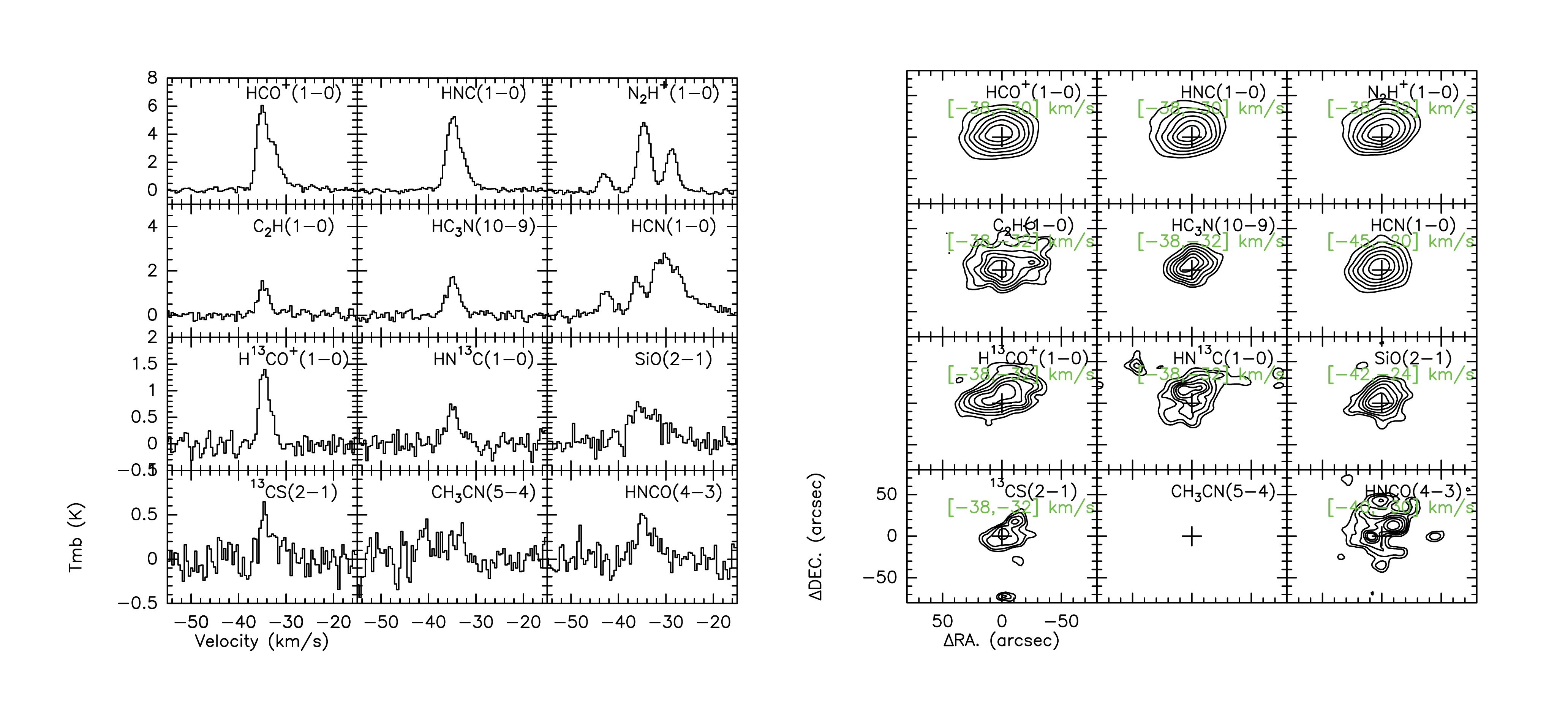,width=6in,height=3in}
 \caption{Molecular spectra
and integrated intensities of AGAL318.948-00.197. Contour levels are
40, 50, . . . , 90 percent of the peak emissions. The black plus
marks the 870 $\mu$m emission peak of AGAL318.948-00.197. The green number show the range of integrations.}
\end{figure}

\begin{figure}
\psfig{file=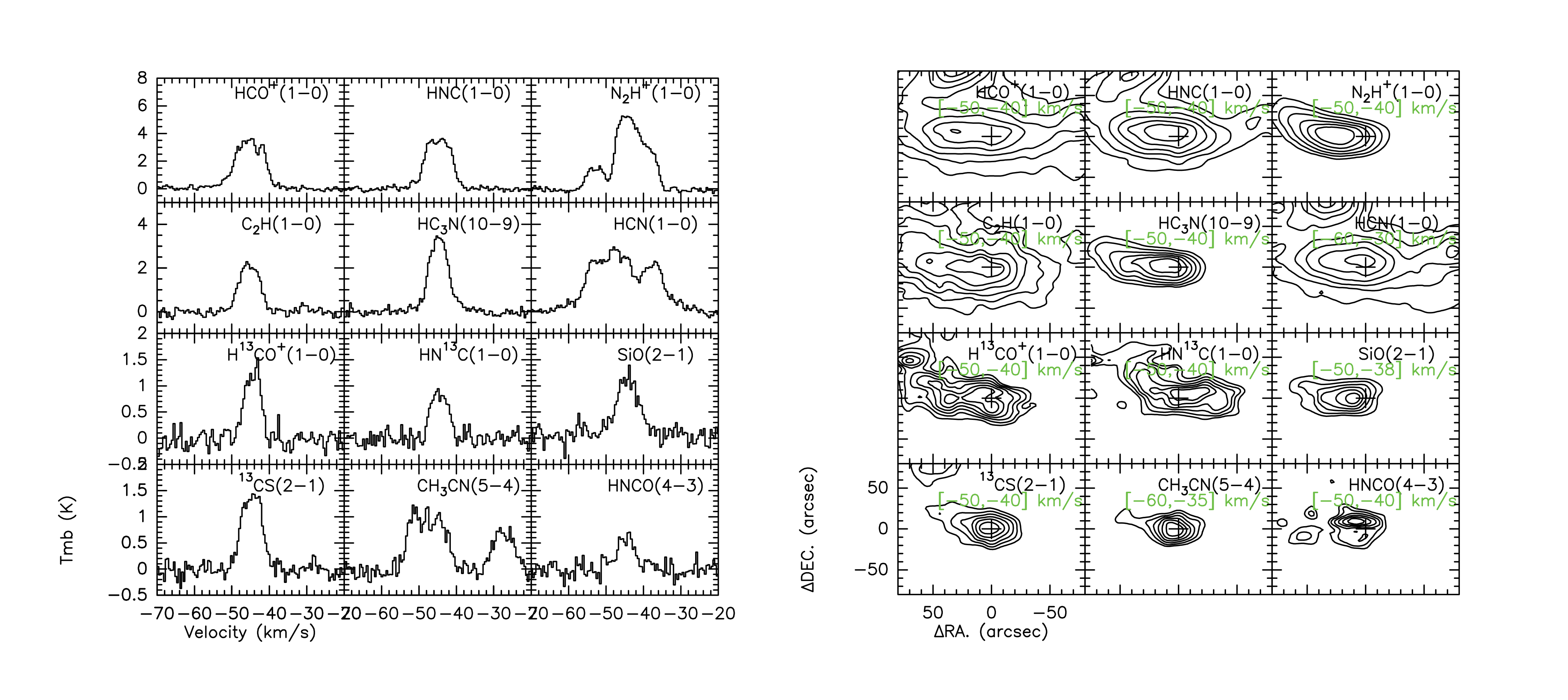,width=6in,height=3in} \caption{Molecular spectra
and integrated intensities of AGAL327.293-00.579. Contour levels are
40, 50, . . . , 90 percent of the peak emissions. The black plus
marks the 870 $\mu$m emission peak of AGAL327.293-00.579. The green number show the range of integrations.}
\end{figure}

\begin{figure}
\psfig{file=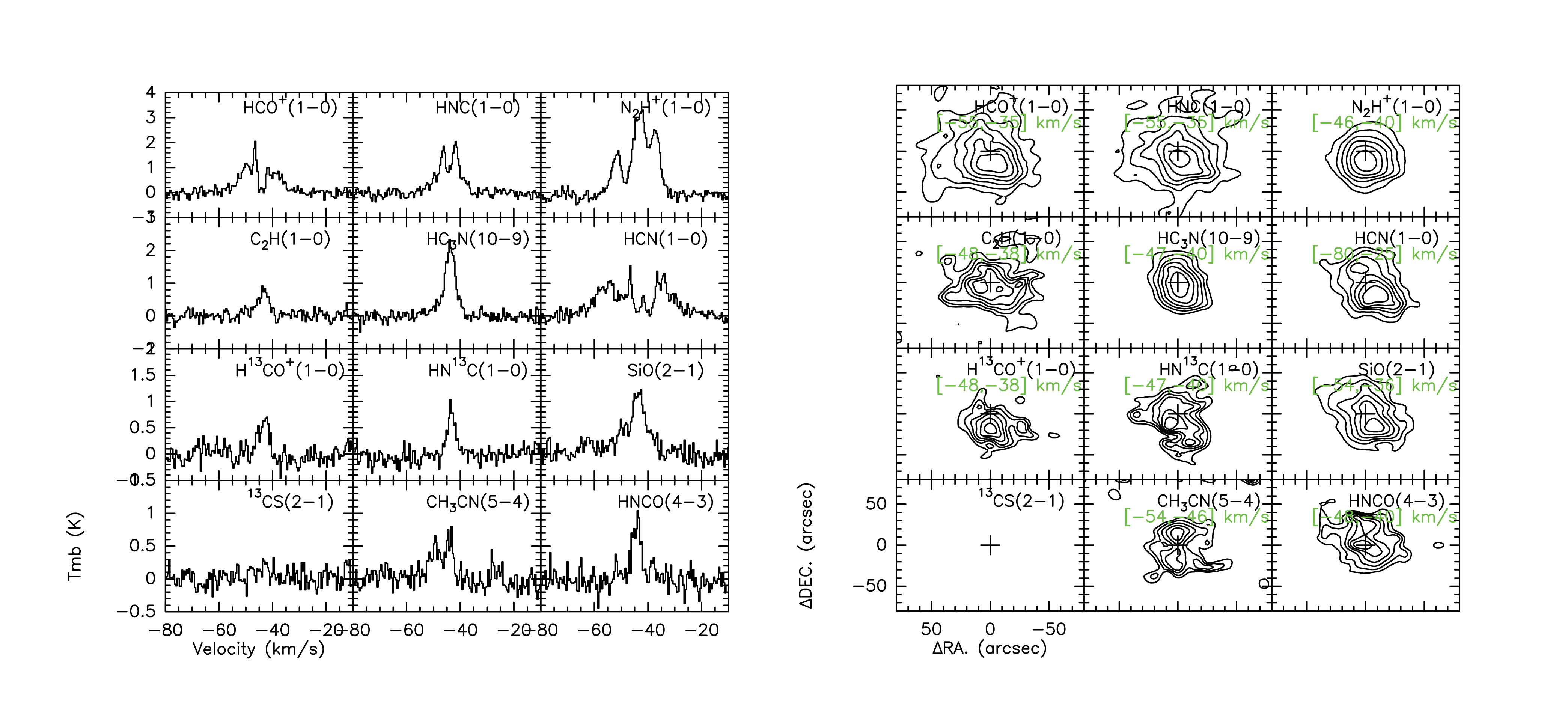,width=6in,height=3in} \caption{Molecular spectra
and integrated intensities of AGAL329.029-00.206. Contour levels are
40, 50, . . . , 90 percent of the peak emissions. The black plus
marks the 870 $\mu$m emission peak of AGAL329.029-00.206. The green number show the range of integrations.}
\end{figure}

\begin{figure}
\psfig{file=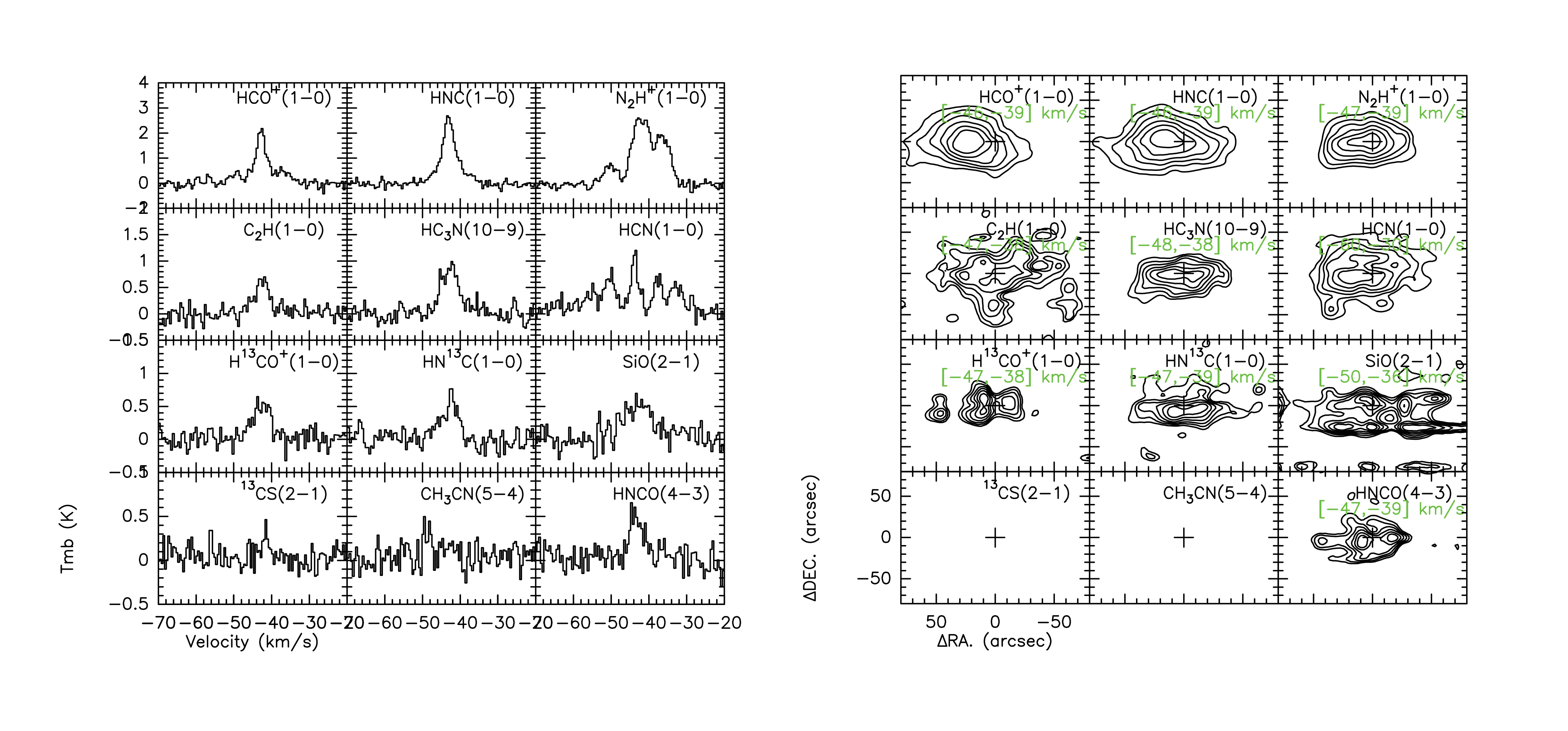,width=6in,height=3in}
 \caption{Molecular spectra
and integrated intensities of AGAL329.066-00.307. Contour levels are
40, 50, . . . , 90 percent of the peak emissions. The black plus
marks the 870 $\mu$m emission peak of AGAL329.066-00.307. The green number show the range of integrations.}
\end{figure}

\begin{figure}
\psfig{file=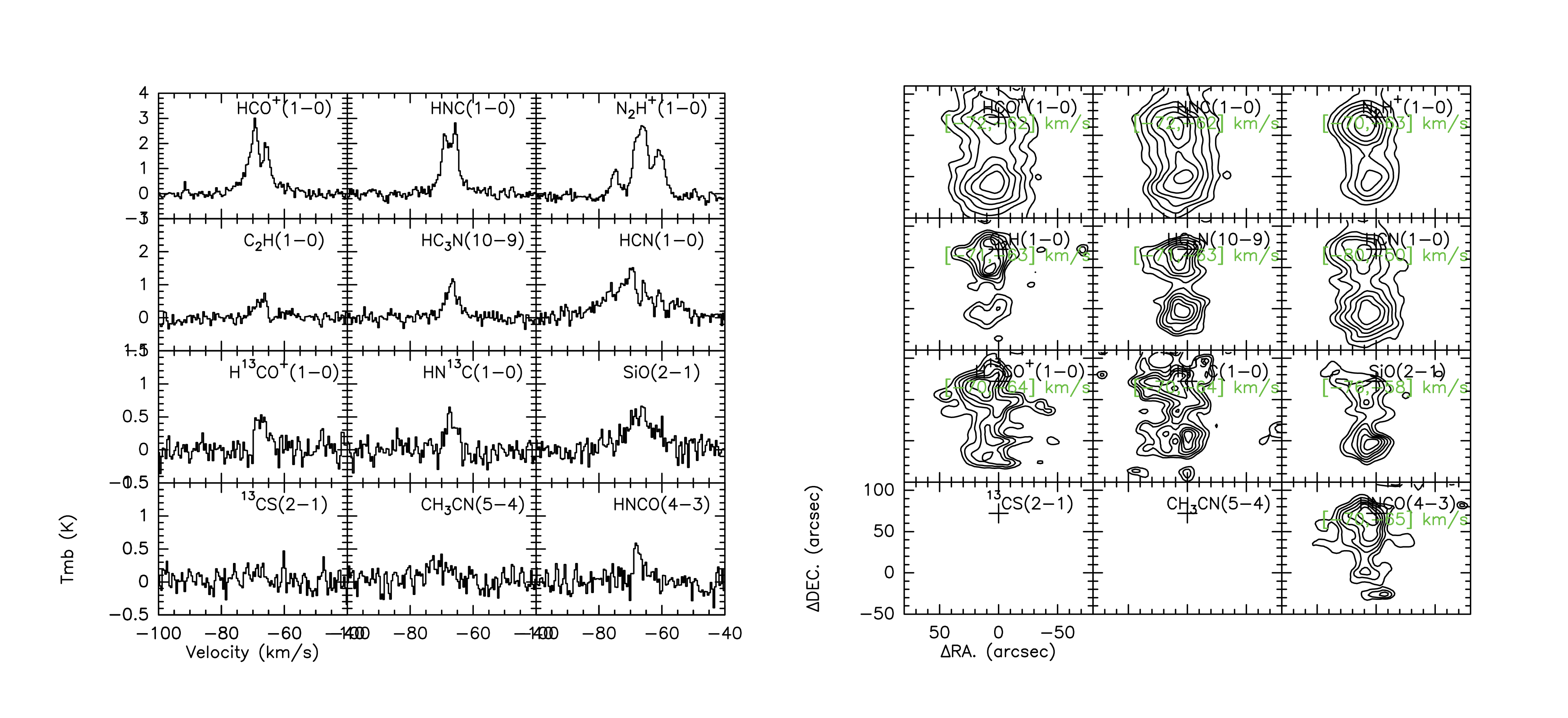,width=6in,height=3in}
 \caption{Molecular spectra
and integrated intensities of AGAL331.709+00.582. Contour levels are
40, 50, . . . , 90 percent of the peak emissions. The black plus
marks the 870 $\mu$m emission peak of AGAL331.709+00.582. The green number show the range of integrations.}
\end{figure}

\begin{figure}
\psfig{file=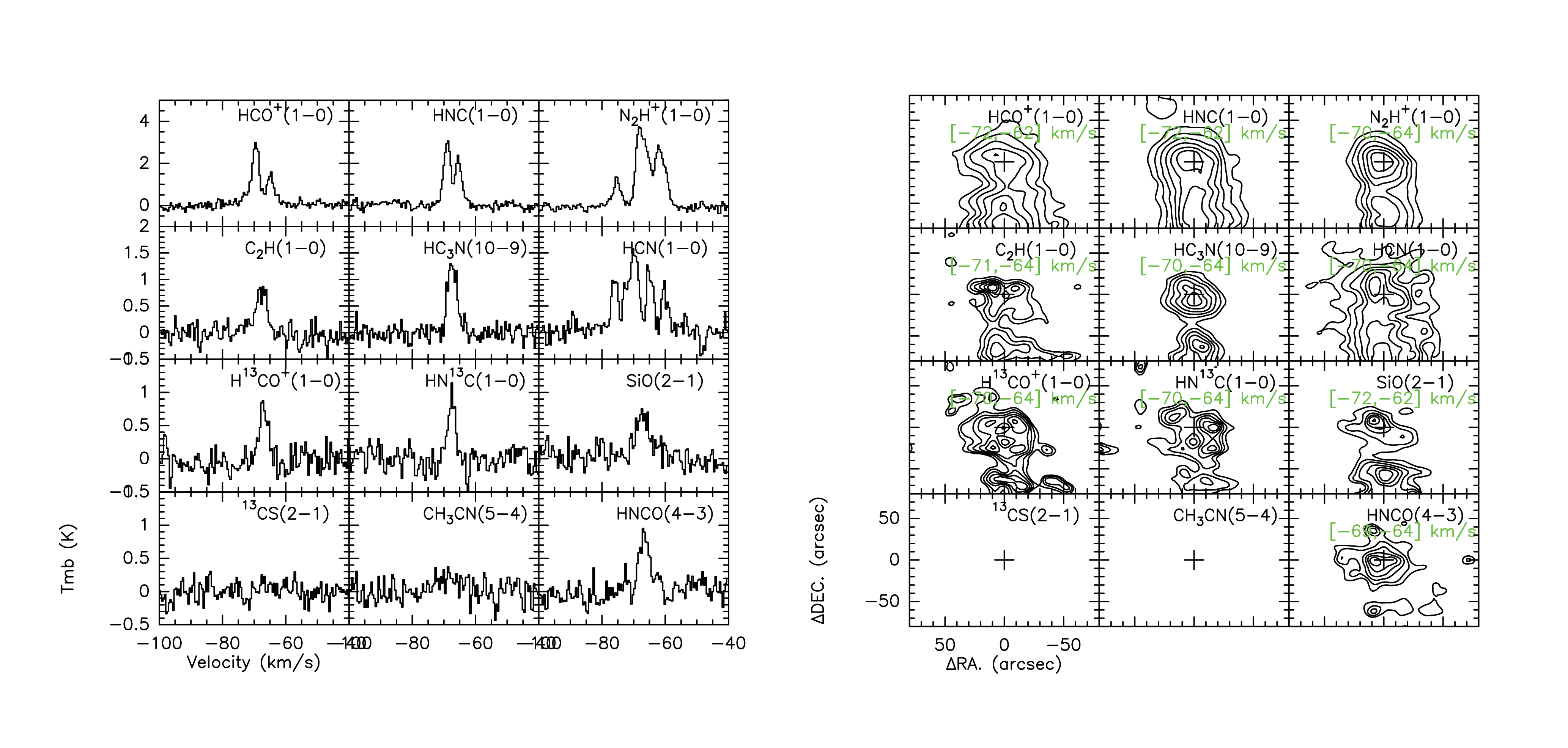,width=6in,height=3in} \caption{Molecular spectra
and integrated intensities of AGAL331.709+00.602. Contour levels are
40, 50, . . . , 90 percent of the peak emissions. The black plus
marks the 870 $\mu$m emission peak of AGAL331.709+00.602. The green number show the range of integrations.}
\end{figure}

\begin{figure}
\psfig{file=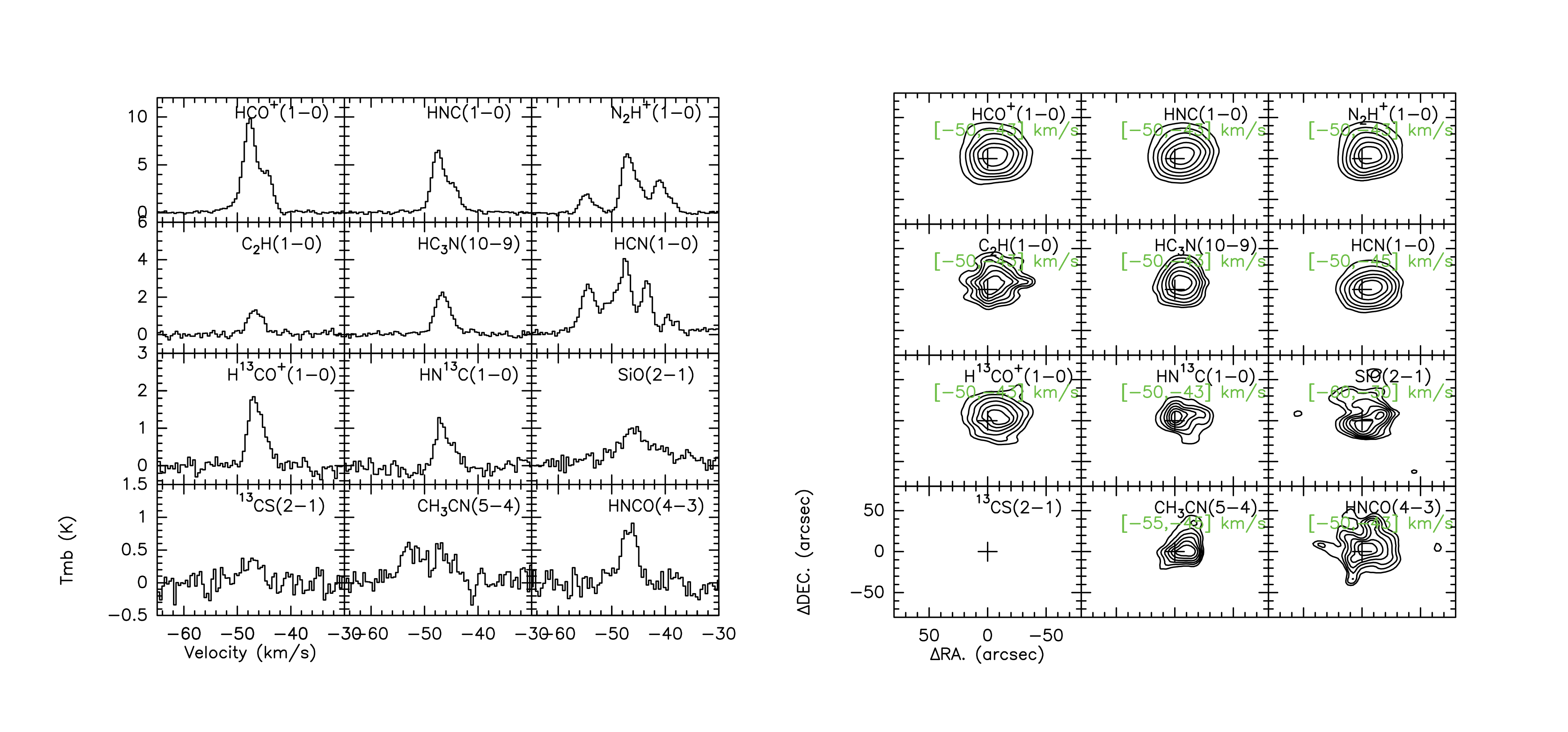,width=6in,height=3in} \caption{Molecular spectra
and integrated intensities of AGAL335.586-00.291. Contour levels are
40, 50, . . . , 90 percent of the peak emissions. The black plus
marks the 870 $\mu$m emission peak of AGAL335.586-00.291. The green number show the range of integrations.}
\end{figure}

\begin{figure}
\psfig{file=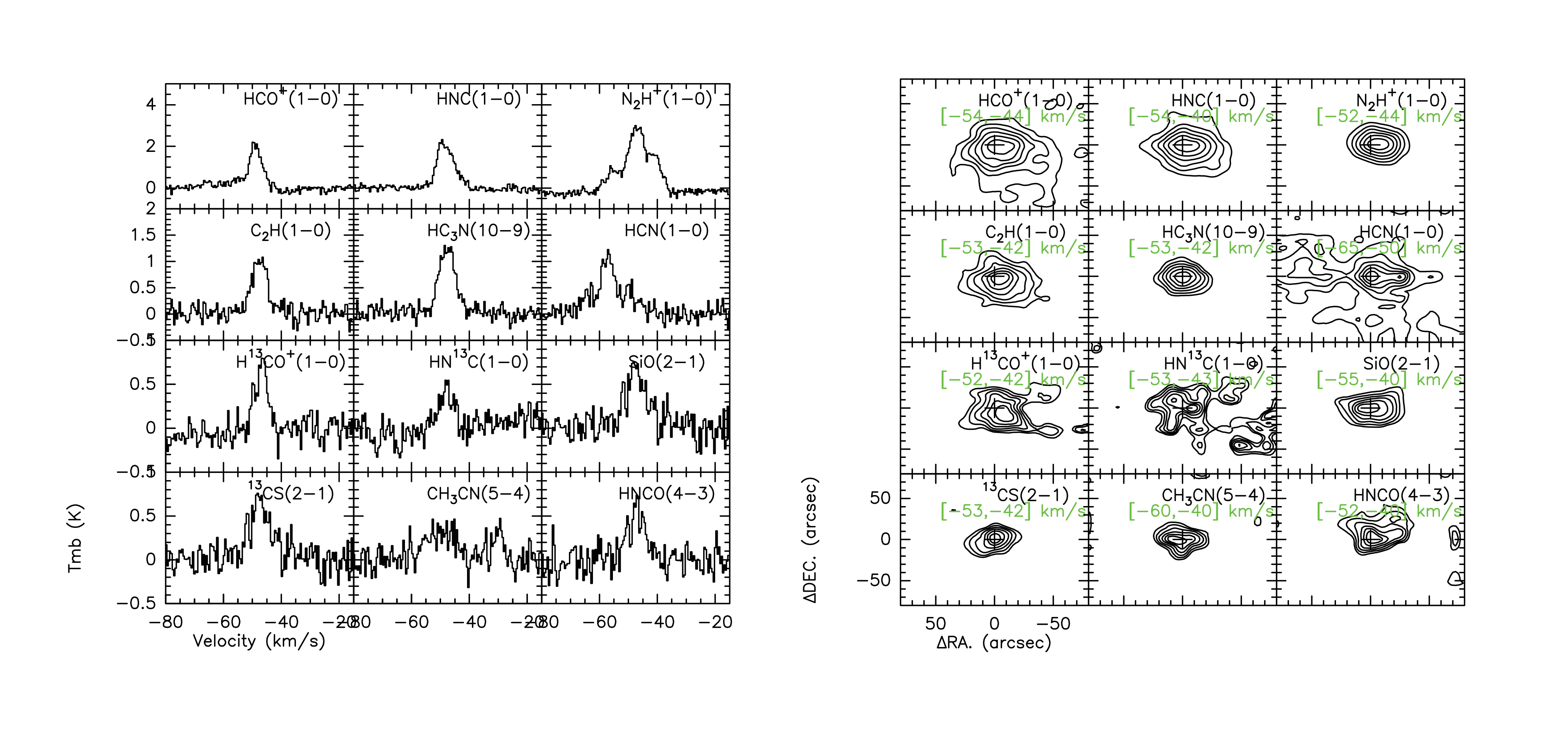,width=6in,height=3in}
 \caption{Molecular spectra
and integrated intensities of AGAL337.704-00.054. Contour levels are
40, 50, . . . , 90 percent of the peak emissions. The black plus
marks the 870 $\mu$m emission peak of AGAL337.704-00.054. The green number show the range of integrations.}
\end{figure}

\begin{figure}
\psfig{file=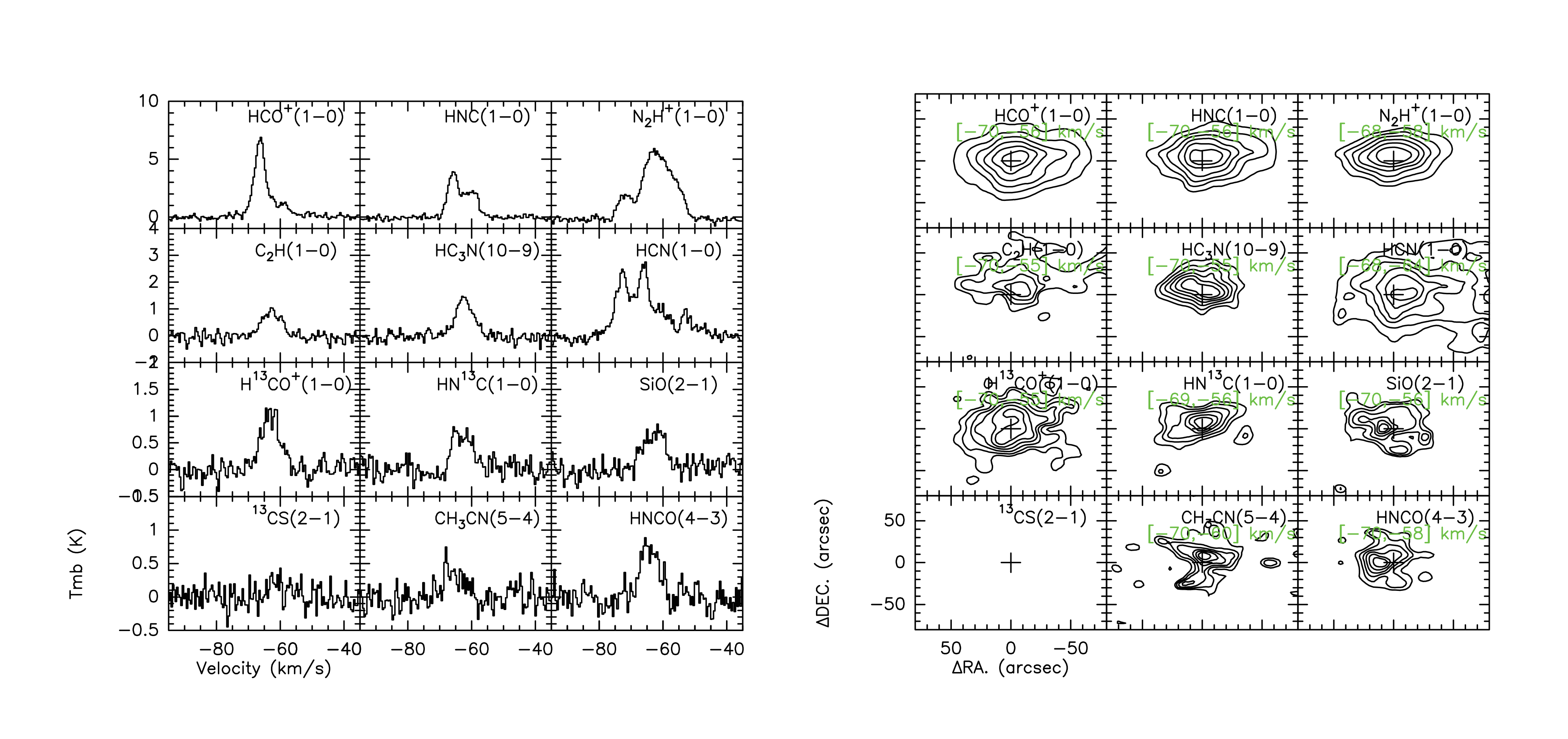,width=6in,height=3in}
 \caption{Molecular spectra
and integrated intensities of AGAL338.926+00.554. Contour levels are
40, 50, . . . , 90 percent of the peak emissions. The black plus
marks the 870 $\mu$m emission peak of AGAL338.926+00.554. The green number show the range of integrations.}
\end{figure}

\begin{figure}
\psfig{file=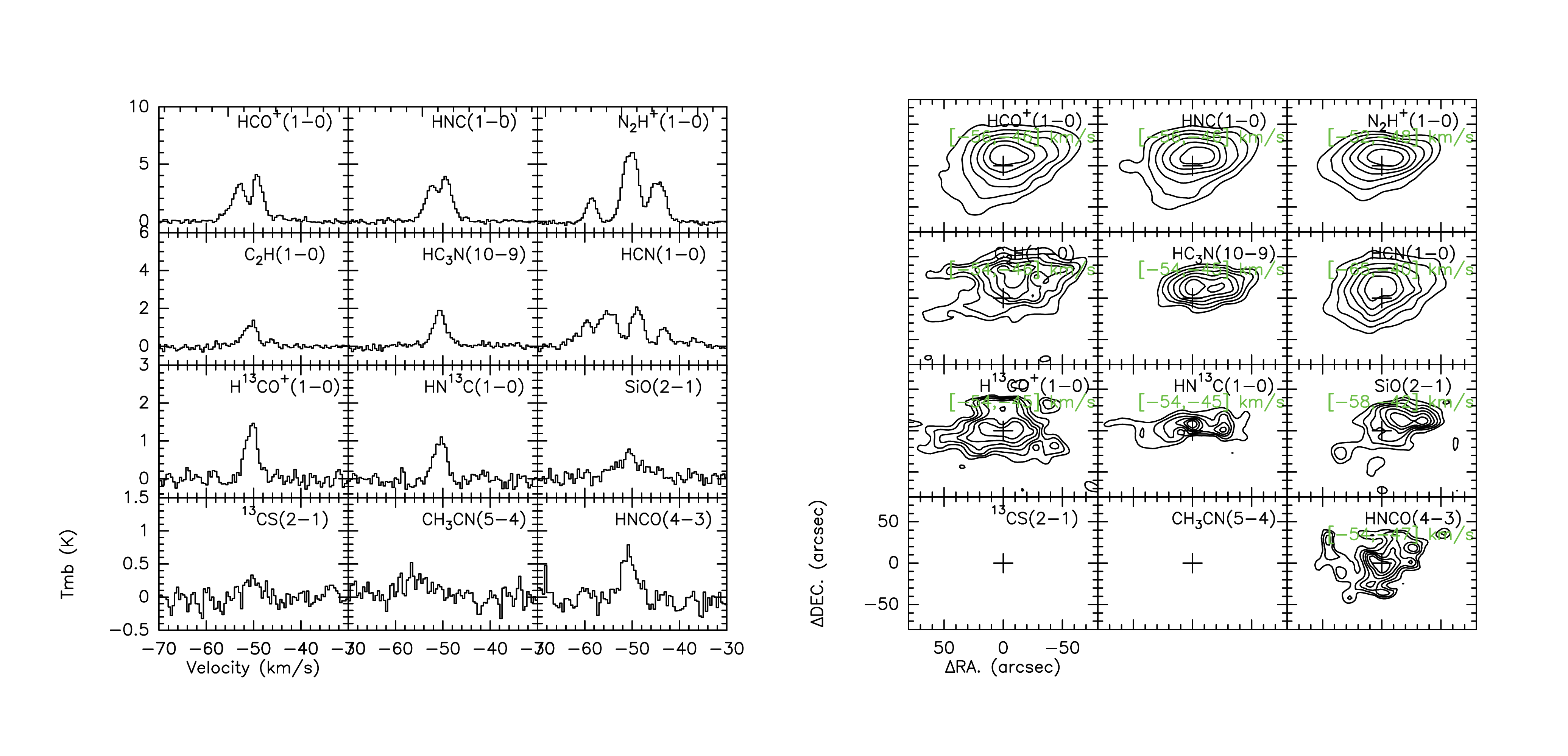,width=6in,height=3in}
 \caption{Molecular spectra
and integrated intensities of AGAL340.248-00.374. Contour levels are
40, 50, . . . , 90 percent of the peak emissions. The black plus
marks the 870 $\mu$m emission peak of AGAL340.248-00.374. The green number show the range of integrations.}
\end{figure}

\begin{figure}
\psfig{file=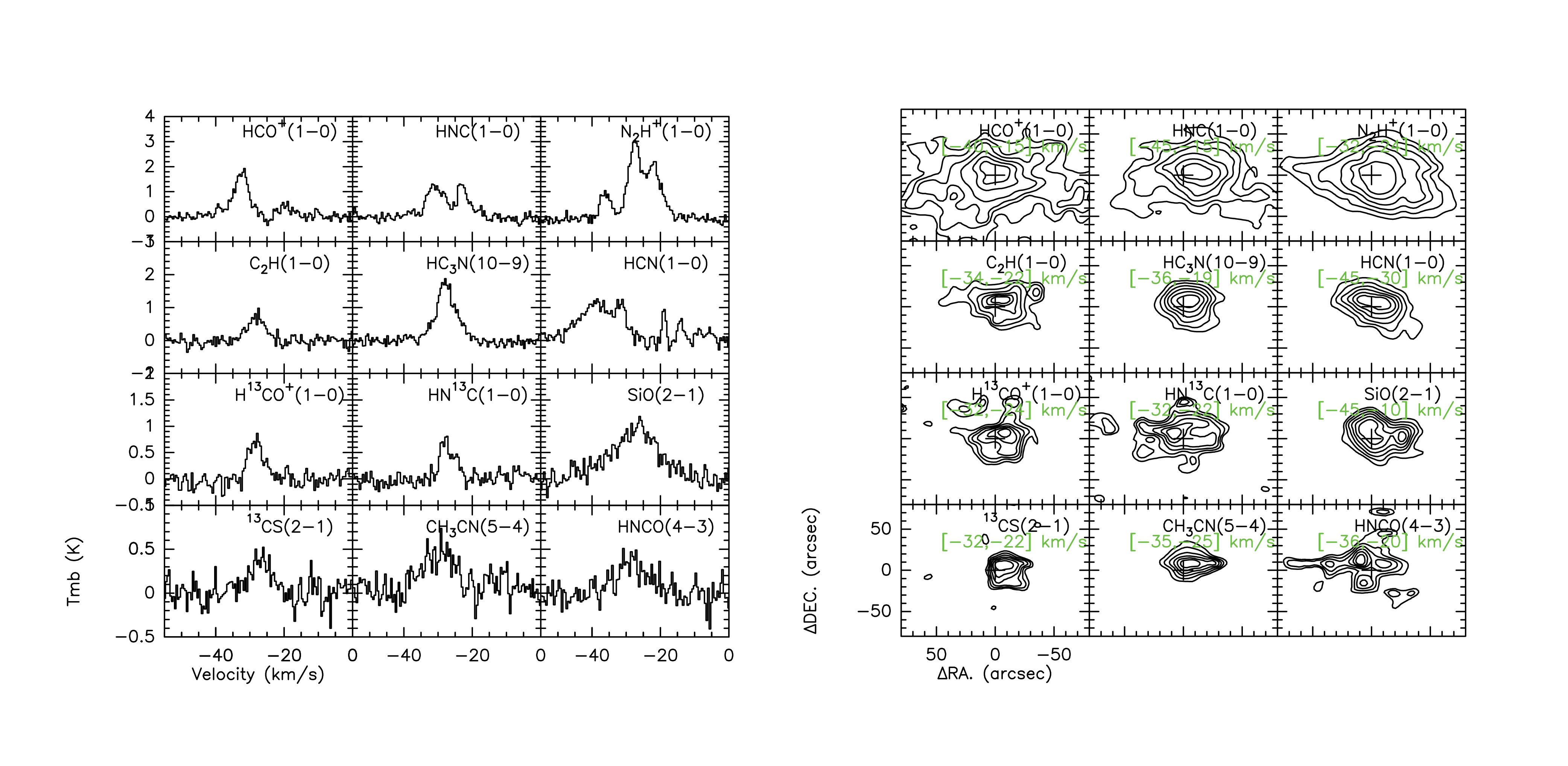,width=6in,height=3in}
 \caption{Molecular spectra
and integrated intensities of AGAL345.003-00.224. Contour levels are
40, 50, . . . , 90 percent of the peak emissions. The black plus
marks the 870 $\mu$m emission peak of AGAL345.003-00.224. The green number show the range of integrations.}
\end{figure}

\begin{figure}
\psfig{file=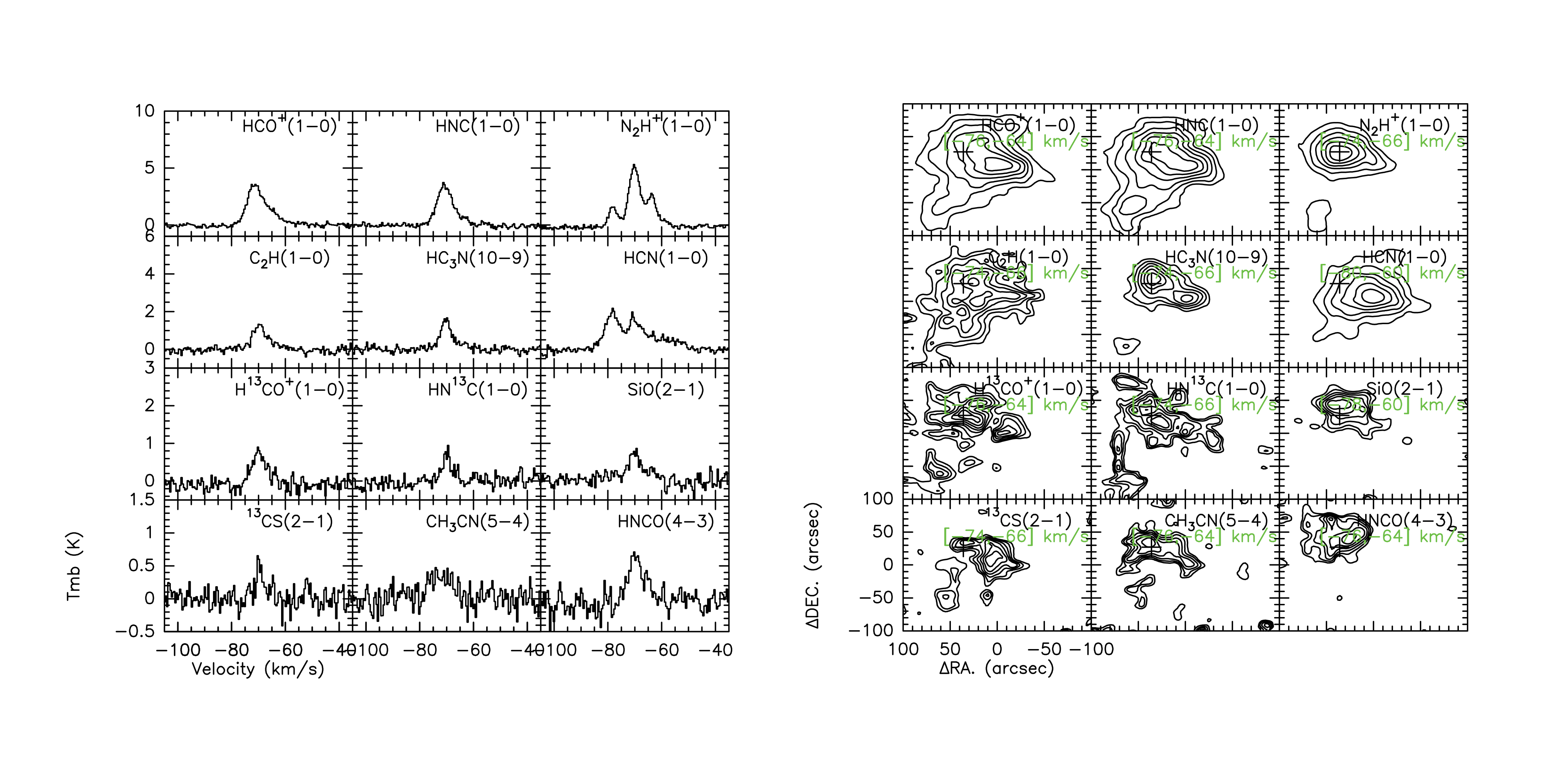,width=6in,height=3in}
 \caption{Molecular spectra
and integrated intensities of AGAL350.111+00.089. Contour levels are
40, 50, . . . , 90 percent of the peak emissions. The black plus
marks the 870 $\mu$m emission peak of AGAL350.111+00.089. The green number show the range of integrations.}
\end{figure}

\begin{figure}
\psfig{file=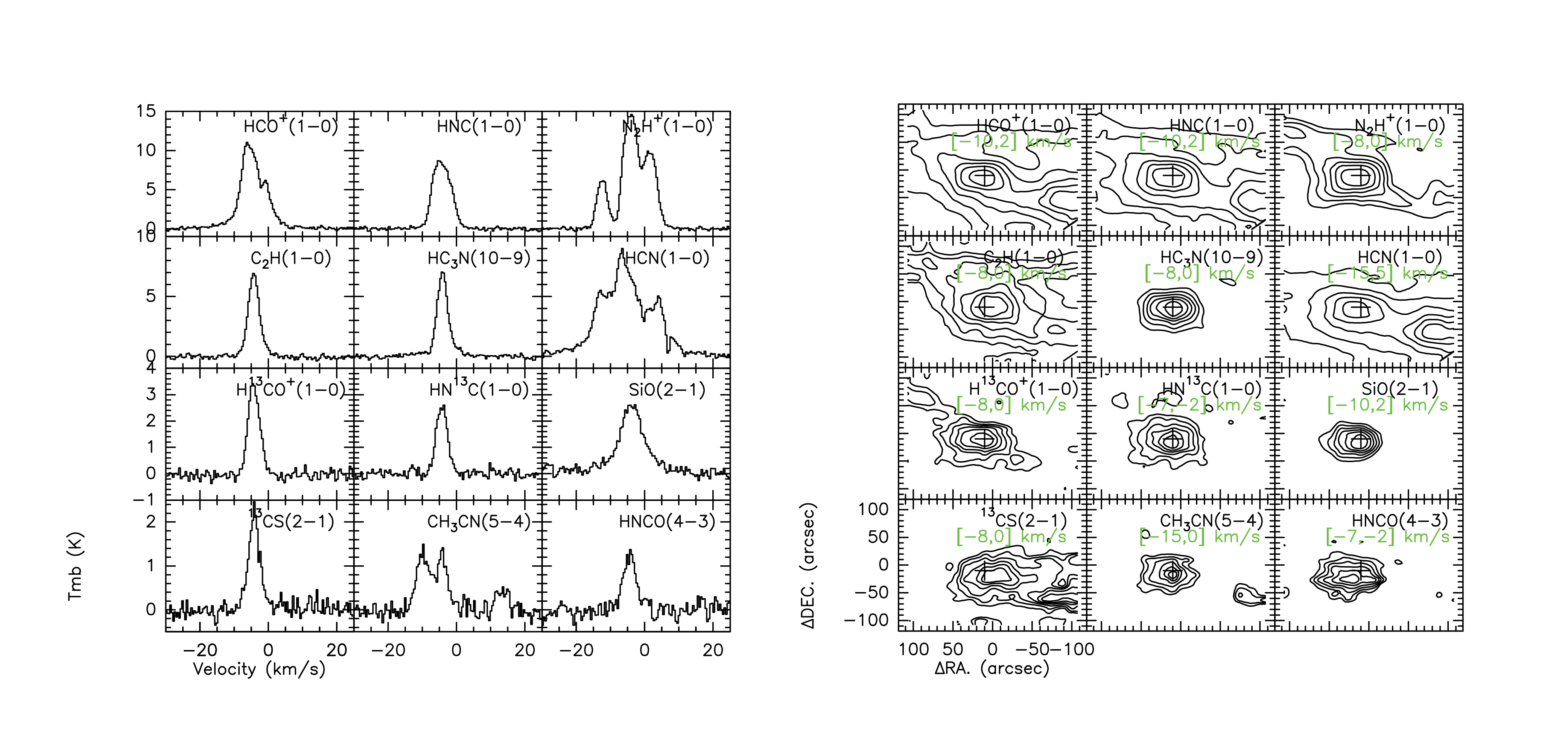,width=6in,height=3in}
 \caption{Molecular spectra
and integrated intensities of AGAL351.444+00.659. Contour levels are
40, 50, . . . , 90 percent of the peak emissions. The black plus
marks the 870 $\mu$m emission peak of AGAL351.444+00.659. The green number show the range of integrations.}
\end{figure}

\begin{figure}
\psfig{file=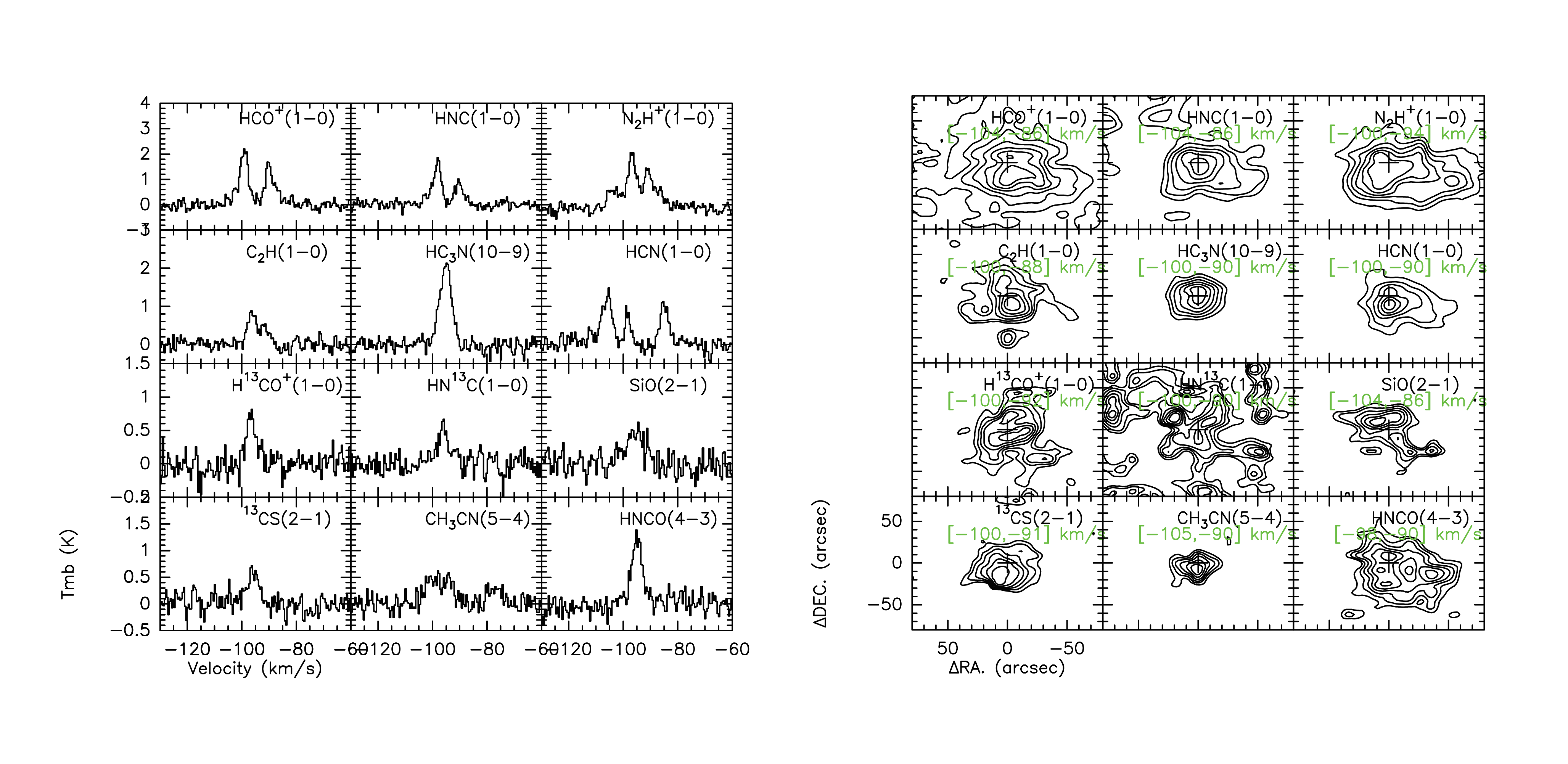,width=6in,height=3in}
 \caption{Molecular spectra
and integrated intensities of AGAL351.581-00.352. Contour levels are
40, 50, . . . , 90 percent of the peak emissions. The black plus
marks the 870 $\mu$m emission peak of AGAL351.581-00.352. The green number show the range of integrations.}
\end{figure}

\begin{figure}
\psfig{file=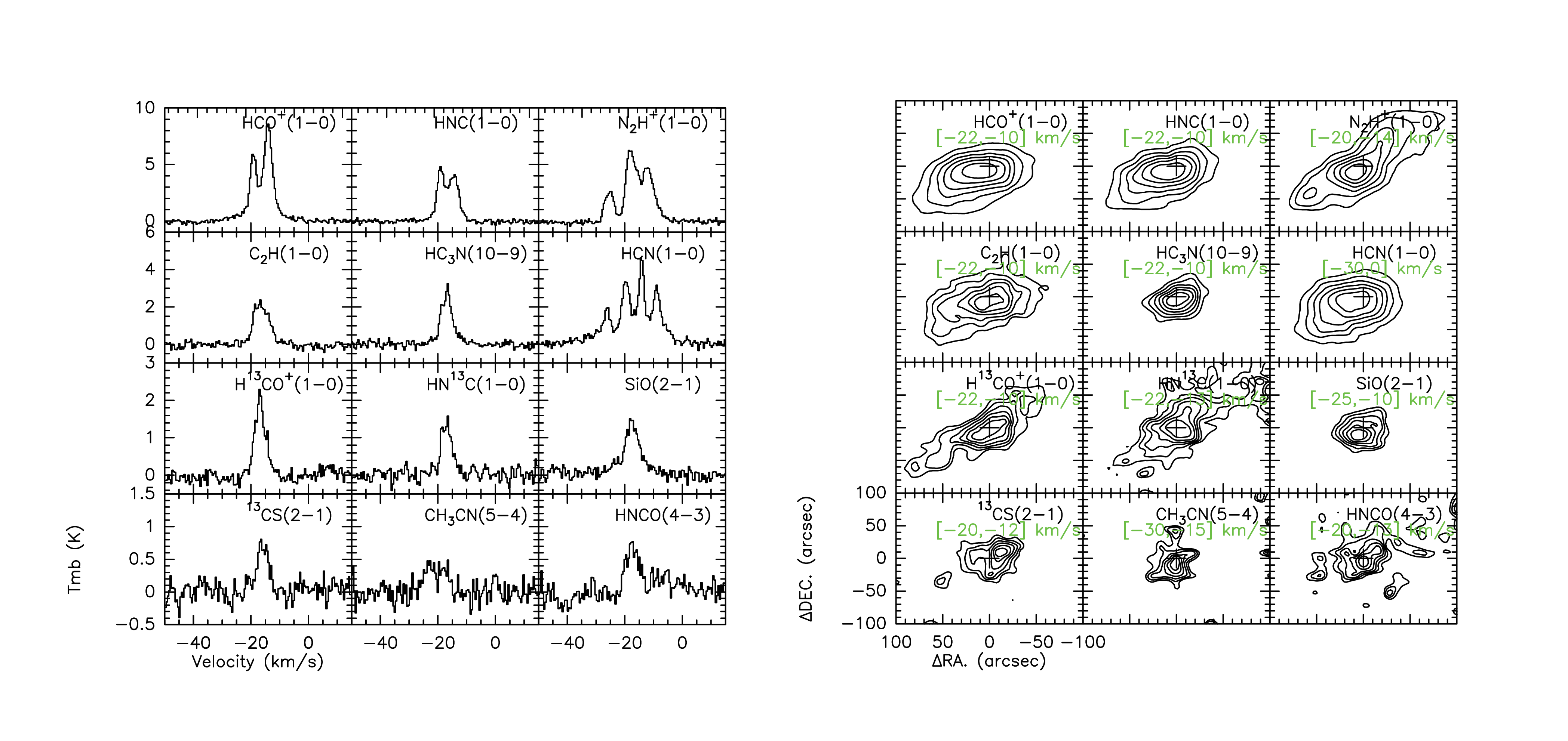,width=6in,height=3in}
 \caption{Molecular spectra
and integrated intensities of AGAL353.409-00.361. Contour levels are
40, 50, . . . , 90 percent of the peak emissions. The black plus
marks the 870 $\mu$m emission peak of AGAL353.409-00.361. The green number show the range of integrations.}
\end{figure}

\clearpage

\begin{figure}
\psfig{file=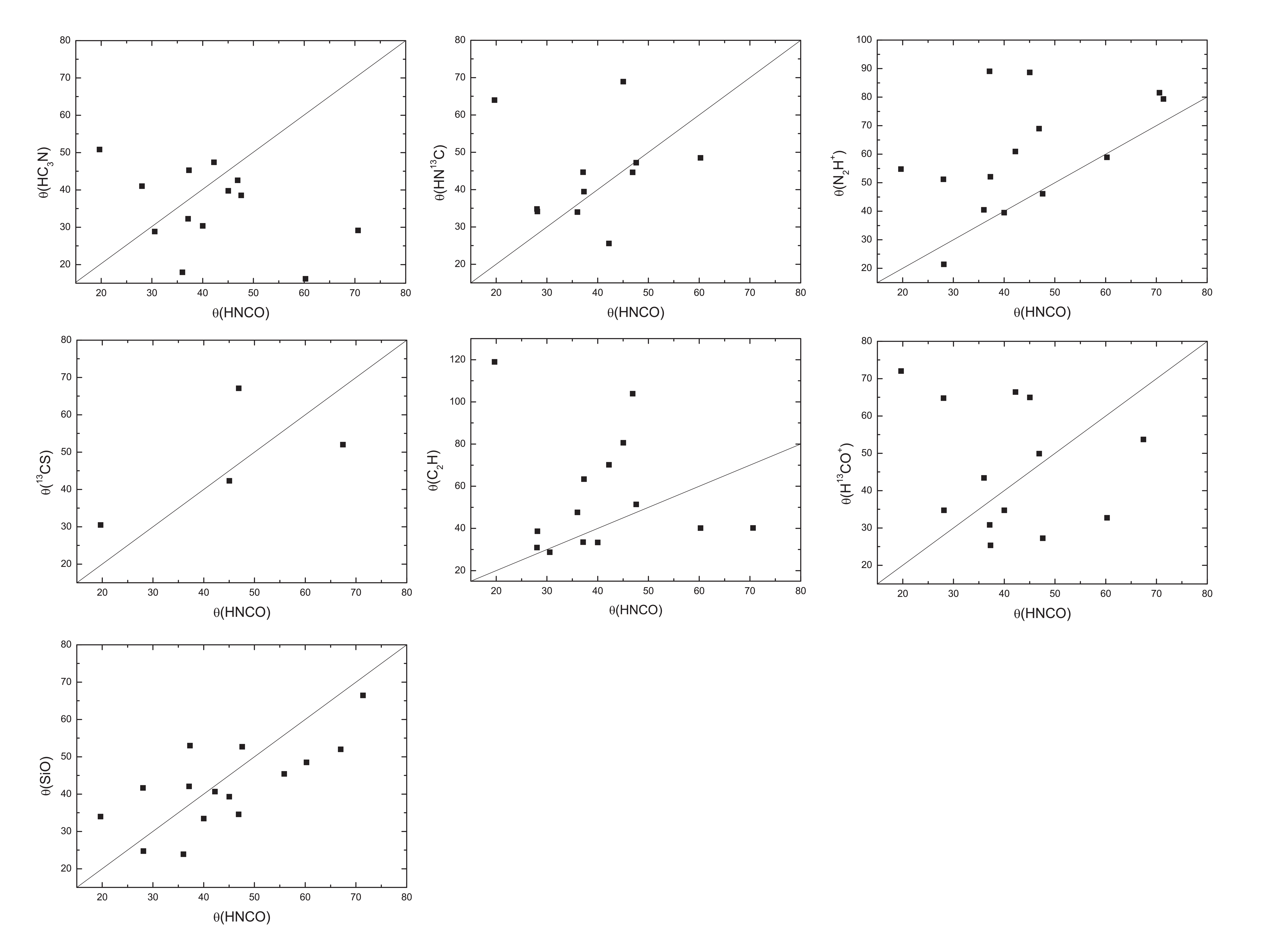,width=6.5in,height=5in} \centering
 \caption{Beam deconvolved angular diameters of HNCO against those of other species. The black lines indicate unity.}
\end{figure}

\begin{figure}
\psfig{file=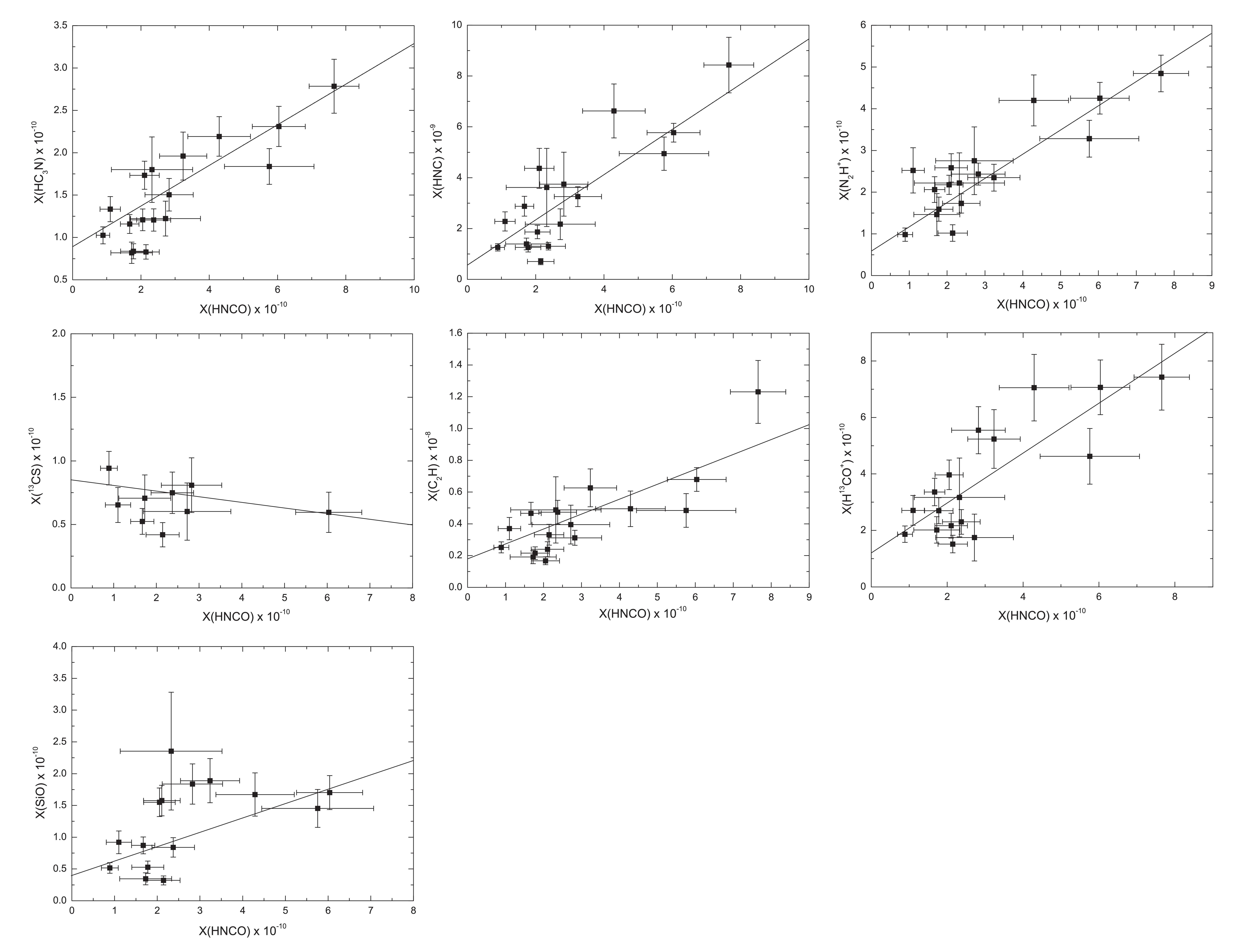,width=6.5in,height=5in}
 \caption{ Plots of the HNCO abundance against those of other species.
The solid line shows the least-squares fit to the data.}
\end{figure}


\begin{figure}
\centering \psfig{file=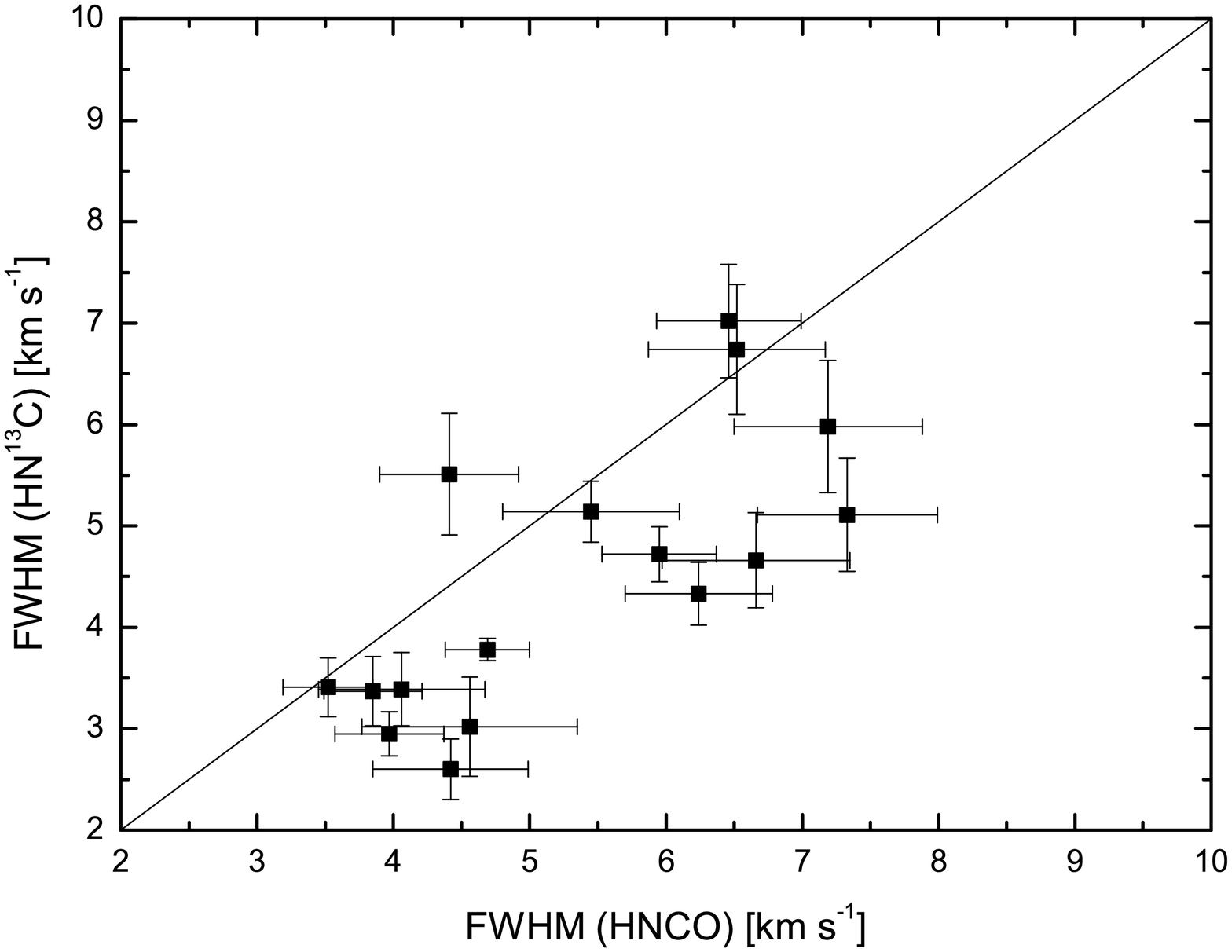,width=2.8in,height=2in} \centering
\psfig{file=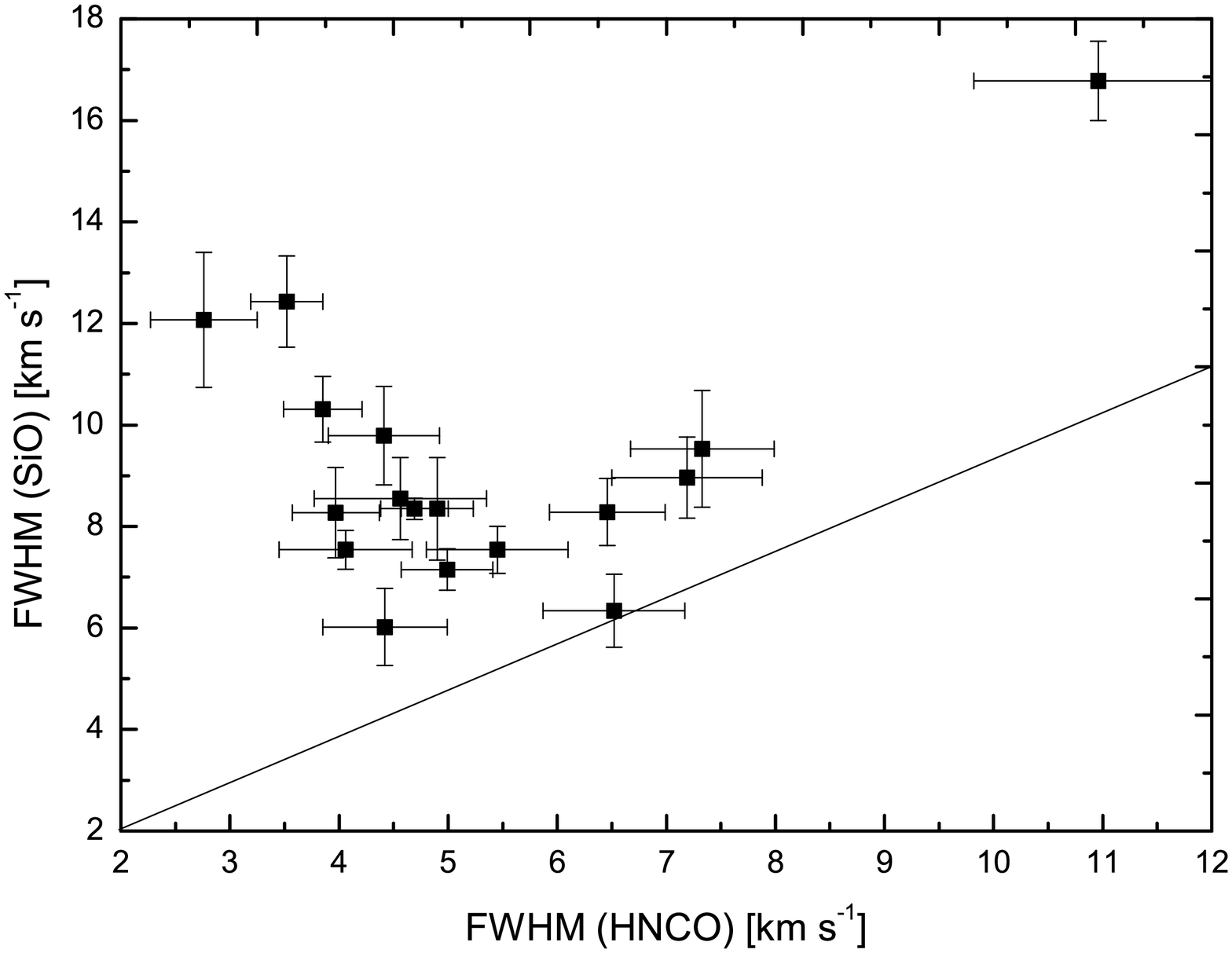,width=2.8in,height=2in} \caption{Plots of the
velocity width of HNCO against that of HN$^{13}$C (the left panle)
and SiO (The right panle). The black lines indicate unity.}
\end{figure}

\begin{figure}
\centering
\psfig{file=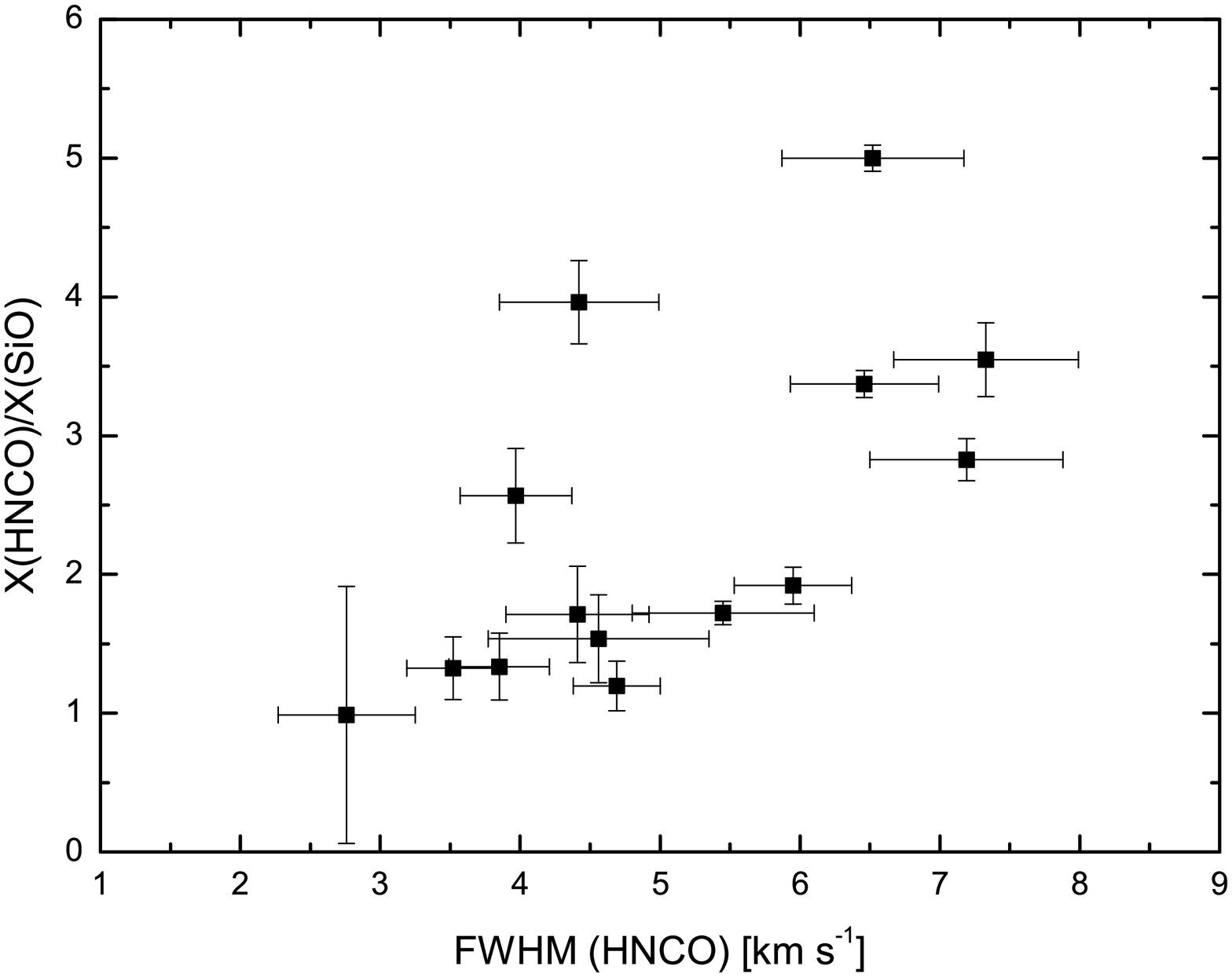,width=2.8in,height=2in}
\psfig{file=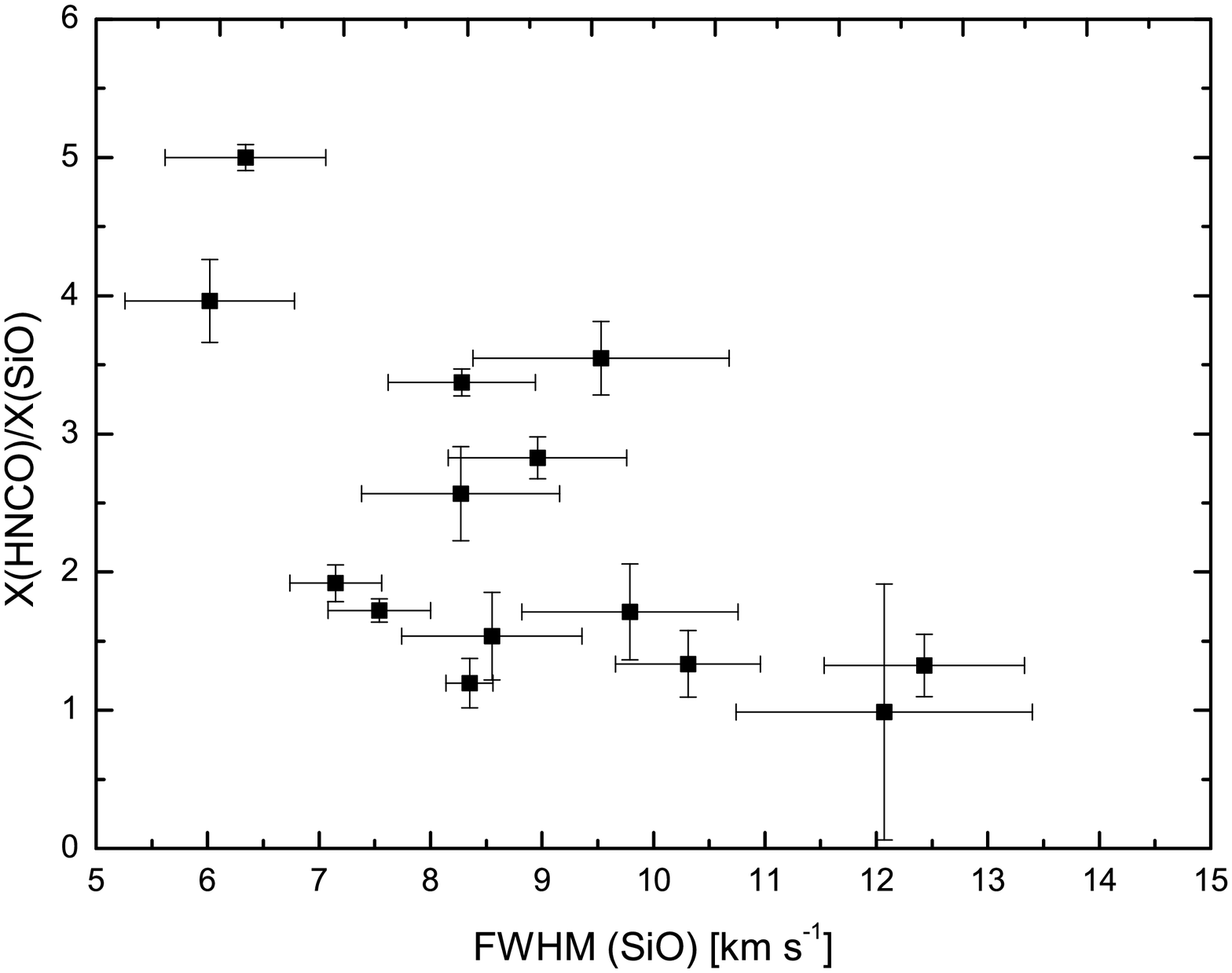,width=2.8in,height=2in}
 \caption{Left: $\chi$(HNCO)/$\chi$(SiO) relative abundance ratio plotted as a function of the HNCO velocity widths.
 Right: $\chi$(HNCO)/$\chi$(SiO) relative abundance ratio plotted as a function of the SiO velocity widths.}
\end{figure}

\begin{figure}
\centering \psfig{file=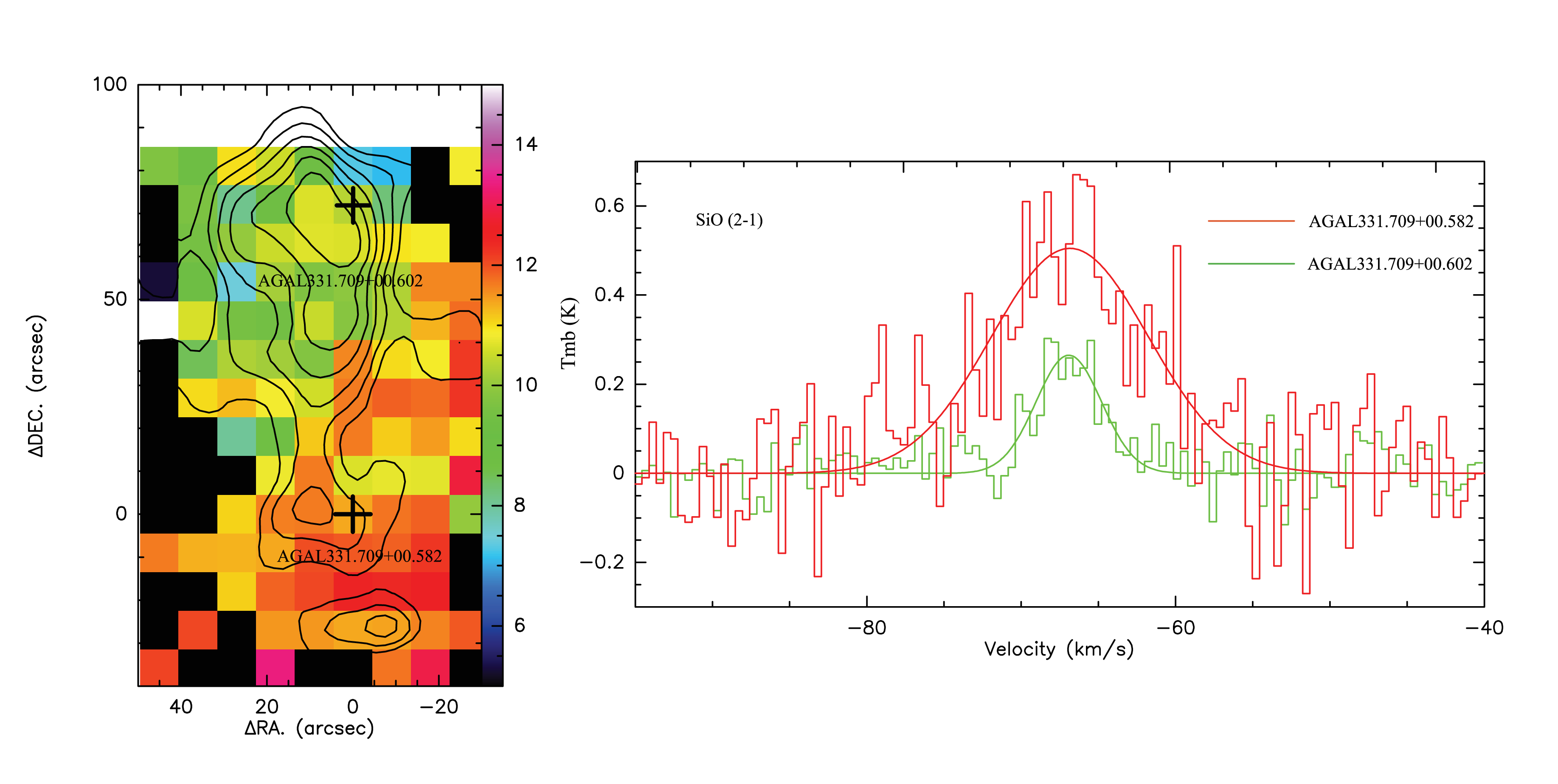,width=6in,height=3in} \caption{Left:
Velocity-width (moment 2) map of SiO overlaid with HNCO integrated
intensity contours (black). The two pluses mark the center locations
of AGAL331.709+00.602 and AGAL331.709+00.582. Right: Spectra of SiO in the center of the two clumps. The red and green
lines are the Gaussian-fitted lines. }
\end{figure}

 \end{document}